\documentclass[twocolumn]{aastex62}
\usepackage{amsmath,amssymb,amstext}
\usepackage{gensymb}
\usepackage{natbib}
\shorttitle{MOPPS I}
\shortauthors{May et al.}

\begin{document}
\title{MOPSS I: Flat Optical Spectra for the Hot Jupiters WASP-4 b and WASP-52b}
\correspondingauthor{E. M. May}
\email{ermay@umich.edu}
\author{E. M. May}
\affil{University of Michigan \\
1085 S. University \\ 
311 West Hall \\ 
Ann Arbor, MI 48109-1107}
\author{M. Zhao}
\affil{The Pennsylvania State University \\
525 Davey Lab \\
University Park, PA 16802}
\author{M. Haidar}
\affil{University of Michigan \\
1085 S. University \\ 
311 West Hall \\ 
Ann Arbor, MI 48109-1107}
\author{E. Rauscher}
\affil{University of Michigan \\
1085 S. University \\ 
311 West Hall \\ 
Ann Arbor, MI 48109-1107}
\author{J. D. Monnier}
\affil{University of Michigan \\
1085 S. University \\ 
311 West Hall \\ 
Ann Arbor, MI 48109-1107}
\begin{abstract}
We present the first results from MOPSS, The Michigan Optical Planetary Spectra Survey, aimed at creating a database of optical planetary transmission spectra all observed, reduced, and analyzed with a uniform method for the benefit of enabling comparative exoplanet studies. We discuss our methods, and present results for our first two targets observed with the Magellan Baade 6.5m telescope, one transit of the Hot Jupiter WASP-4b and two transits of the Hot Saturn WASP-52b. Both targets present flat, featureless spectra, corresponding to the presence of aerosols. We find that the cloud decks must begin no lower than 10$^{-4}$ bar for both planets.  For WASP-52b, we also consider the effects of star spots on the transmission spectrum, including unocculted spots and spots on the stellar limb influencing the light curve limb darkening parameters. We discuss the usefulness of this program in the coming James Webb Space Telescope era.
 \end{abstract}
%

%
\section{Introduction}
The study of exoplanet atmospheres has been a rapidly growing field for many years. Observationally, transmission spectroscopy has grown into a robust way to constrain the composition of a planet's atmosphere, assuming a well-understood host star. The increasing number of planets with observed transmission spectra has moved us towards comparative studies of planetary atmospheres, rather than studies of planets as single, non related, data points (see \cite{Sing2016} for a discussion of the variety of Hot Jupiter atmospheres, \cite{Fu2017} for an updated look at Hubble spectra, and \cite{Crossfield2017} for a look at Neptune-sized planets). With this, comes the need to have sets of uniformly observed and reduced planetary transmission spectra to better make comparisons between planets. In this work, we discuss our first steps towards contributing our own catalog of such planets observed with the Magellan Baade telescope at Las Campanas Observatory in Chile. 
\par Transmission spectroscopy measures the relative size of an exoplanet compared to its host star as a function of wavelength. Because a fraction of the stellar light will filter through the planet's gaseous atmosphere before reaching our observatories, the exact transit depth measured is a combination of the light blocked out by the optically thick portion of the exoplanet and the amount of absorption and scattering of stellar light by the atmosphere. This allows us to reconstruct likely compositions of the atmosphere, including mean molecular weight, molecular and atomic atmospheric constituents, and clouds. In the presence of absorbers or scatterers, the transit depth will appear larger, due to the star light that is removed from our line of sight by the planet's atmosphere. The expected strength of such features scales with the planet's atmospheric scale height; related to the planet's gravity, limb temperature, and mean molecular weight. Therefore, a typical `good' target for transmission spectroscopy is a planet with a high limb temperature, low atmospheric mean molecular weight, or a low gravity. 
\par To date, planet atmospheres are best described as being `cloudy', with a flat and featureless transmission spectrum (uniform transit depth with wavelength); `clear', showing evidence of atomic and molecular absorption; or `hazy', showing evidence of scattering, but few, or diminished, absorption features. The characterization of a planet's atmosphere benefits from the detection or non-detection of this absorption and/or scattering through the atmosphere. At optical wavelengths, the primary source of scattering is Rayleigh scattering. Rayleigh scattering has a wavelength dependence of $\lambda^{-4}$, resulting in a stronger effect at shorter, bluer, wavelengths. These higher amounts of scattering at shorter wavelengths corresponds to a deeper planetary transit, as more light is scattered from our line of sight. Meanwhile the primary source of absorption at optical wavelengths is due to the alkali metals Sodium and Potassium. The presence or lack-there-of of any of these features tells us information about the composition of the planet's atmosphere.
\par In order to best understand the shapes of transmission spectra , we need to leverage multiple types of observatories to combine data from a variety of wavelength regimes. Hubble as been a leading force in infrared transit observations; from small, Neptune-like planets \citep[e.g.][]{Ehrenreich2014, Kreidberg2014}; to large, Hot Jupiters \citep[e.g.][]{Vidal2003,Sing2008}. More recently, \cite{Tsiaras2017} did a re-analysis of the Hubble observations for 30 exoplanets in order to provide a uniformly reduced and analyzed sample of planets. At optical wavelengths, numerous ground-based observatories are working on surveys; including the ACCESS group, also at the Magellan telescopes at Las Campanas, with results for GJ 1214 b \citep[]{Rackham2017}; and the Gran Telescopio Canarias exoplanet transit spectroscopy survey, currently with results for numerous planets \citep[see][]{GTC0,GTC1,GTC2,GTC3,GTC4,GTC5,Chen2017,GTC7,GTC8}. Because telescope time is limited, it is important to make use of all resources available to us to study the atmospheres of new planets, as well as ensure reproducibility of results across telescopes, instruments, and reduction methods. 
\par In this work, we present the first results of the Michigan Optical Planetary Spectra Survey (MOPSS) at the 6.5-meter Magellan Baade Telescope. This survey is designed to create a catalog of uniformly observed and reduced transmission spectra for the benefit of enabling comparative exoplanet studies. WASP-4b and WASP-52b provide two case studies with which we built our complete reduction and analysis pipeline. Both planets were selected for their size, expected transmission signals, and/or ease of observations from the Magellan Baade Telescope.
\par In Section \ref{observation}, we discuss the instrument used, our observational set-up, and previous observations of our targets. In Section \ref{analysis}, we outline the steps in our data reduction pipeline, as well as the steps we take in the analysis of the resulting light curves. Section \ref{results} details our resulting transmission spectra for both targets, and discusses the affects of star spots on our transits. Section \ref{future} discusses the future steps and goals for MOPSS, as well as the benefit of ground based optical planetary studies in the coming James Webb Space Telescope (JWST) era.

\section{Observations} \label{observation}

\subsection{The IMACS Instrument} 
IMACS (Inamori-Magellan Areal Camera \& Spectrograph) is a wide-field imager and optical multi-object spectrograph located on the 6.5-meter Magellan Baade Telescope at Las Campanas Observatory. In this work, we use the f/2 camera on IMACS in spectroscopic mode while simultaneously observing our target and a number of calibration stars with custom-made observing masks. The IMACS f/2 CCD consists of 8 separate rectangular chips in a 4x2 arrangement. There is a gap of $\sim$57 pixels between the short edges of the chips, and $\sim$92 pixels between the long edges of the chips. 
\par
\begin{figure}
	\epsscale{1.2}
	\plotone{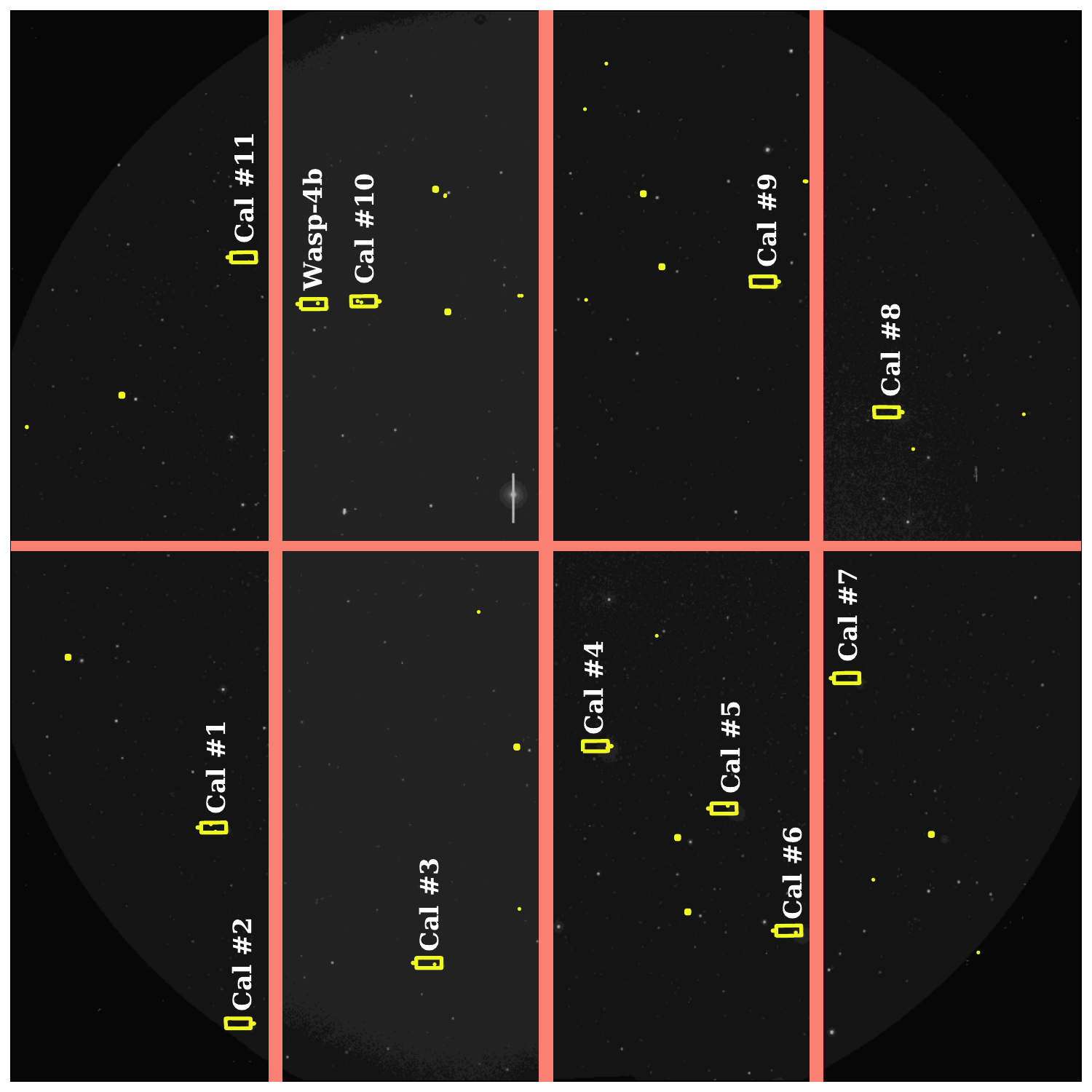}
	\hspace{0.1cm}
	\plotone{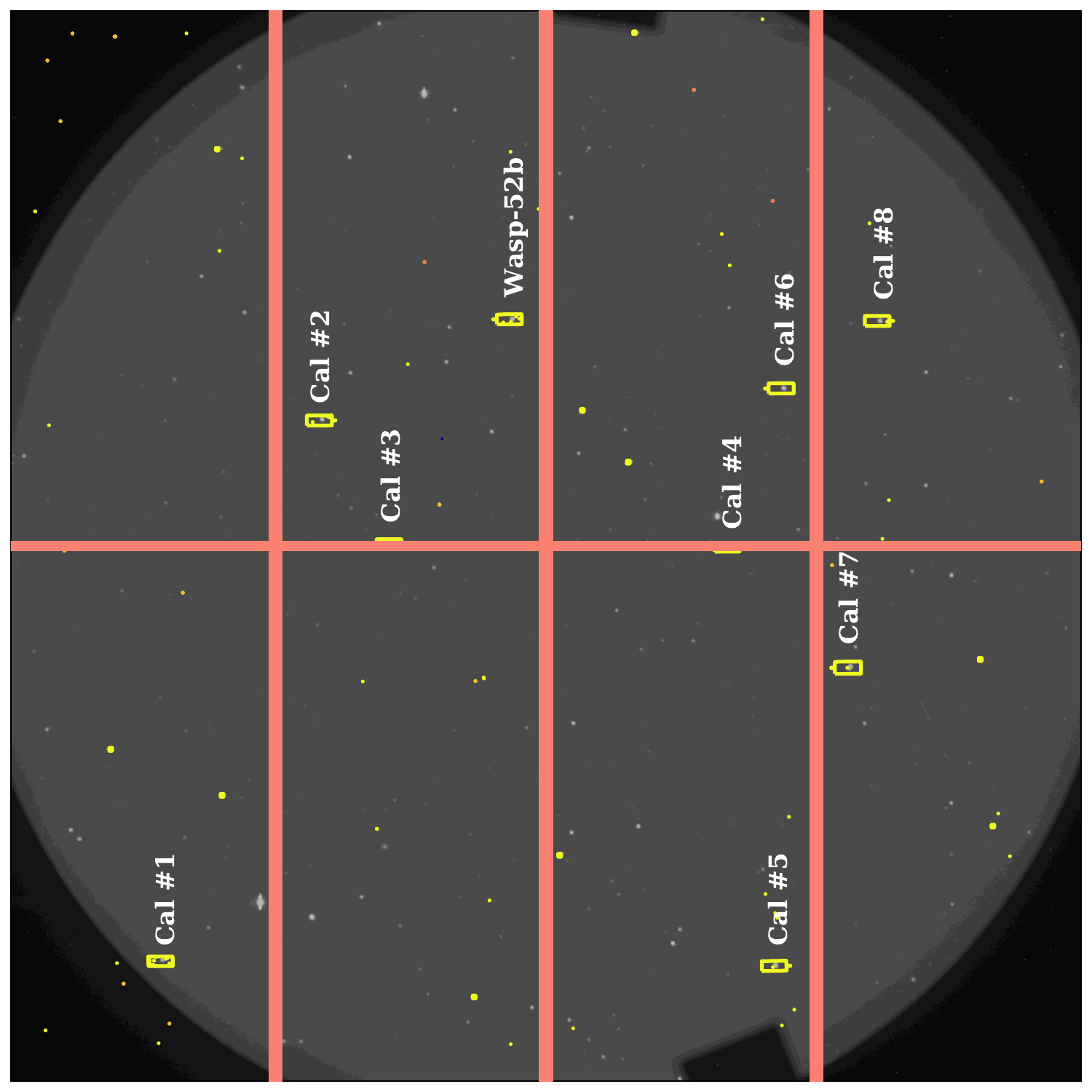} 
	\caption{Field of Views for WASP-4b (top) and WASP-52b (bottom) with calibrator stars marked. The large yellow boxes represent the size of the slits cut on the instrument masks, the small yellow marks are alignment stars for the mask (not used for data collection).} \label{FOVS}
\end{figure}

 For both targets discussed in this work, we use the 300 lines/mm grism at a blaze angle of 17.5$^{\degree}$, resulting in theoretical wavelength coverage from 3900\AA{ }to 8000\AA, but practically, from $\sim$4600\AA{ } to $\sim$8000\AA{ } due to wavelength calibration, sources of noise (O$_2$ absorption in our atmosphere), and second order contamination beyond 8000\AA{ }(future work will use a blocking filter designed to push this contamination to longer wavelengths). Our instrument masks are designed to maximize the number of calibrator stars with complete wavelength coverage while keeping the main target on a central chip (see Figure \ref{FOVS}). Calibrator stars were chosen based on their magnitudes and spectral type, or their color if spectral type was not available. Ideal calibrators are the same spectral type or those with the most similar B-V and J-K colors, with a requirement that the star be within $\sim$ 0.75 magnitudes of the main target. For each object, we used wide slits (16'') with large lengths (20'') to minimize slit losses and improve fits for background subtraction. With this setup, our resolution is limited by the seeing each night (see Sections \ref{W4_obs} and \ref{W52_obs} for seeing data for each night). Our observational efficiency was limited by the read-out time for the CCD; at 81 seconds in fast mode and 1x1 binning, which is comparable to the exposure time for each frame for these targets.

\subsection{The planet WASP-4b} \label{W4_obs}
WASP-4b is a $\sim$1600K Hot Jupiter orbiting a G-type star. WASP-4b has previously been targeted for transmission spectroscopy by Hubble, with both transit and secondary eclipse measured \citep[See][]{Ranjan2014}. \cite{Ranjan2014} find that their transmission measurements heavily depend on their reduction methods due to chip saturation, and as such, do not report results. Recent work by \cite{Huitson2017} presents complete optical transmission spectra for WASP-4b observed at the Gemini South telescope from 4400-9400\AA. We compare their results to ours in Section \ref{w4_results}.
\par WASP-4b was observed on the night of August 19th, 2015 with an exposure time of 90 seconds and a total of 112 exposures, for an observational efficiency of 53\%. Thin clouds passed during the night, with increased cloud coverage toward the end of the night. We discard the final 15 exposures due to these clouds, leaving us with 97 science frames. Seeing ranged from 1'' to 1.3'' during the course of the night, with the object's airmass ranging from $\sim$1.0 to $\sim$1.5. Our wide slits mean that we are not concerned with slit losses even during these unfavorable seeing conditions. In addition to our main target, our instrument mask allowed us to simultaneously target 11 calibrator stars (see Table \ref{w4cals}). The large number of calibrator stars observed allows us to be highly selective when choosing those to use for flux calibration, as well as reserve one calibrator to serve as a `check' star. Since these calibrators should be constant throughout the night, this `check' star provides a method to ensure our pipeline is correctly accounting for airmass and seeing effects throughout the night. 

\begin{figure}
	\epsscale{1.2}
	\plotone{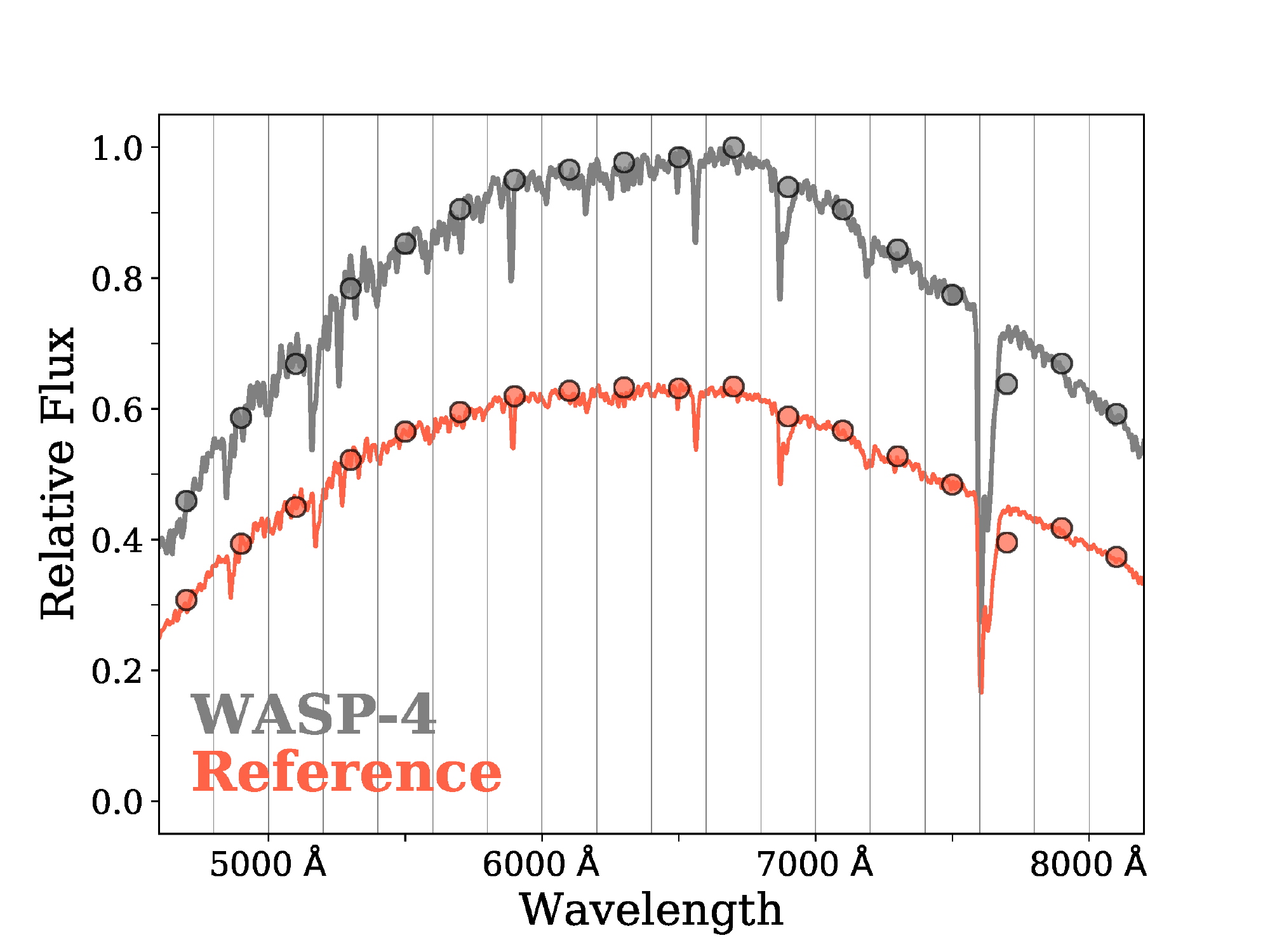}
	\caption{Spectra for WASP-4 (grey line) and the representative calibrator (red line) from 4600 \AA through 8200 \AA. The edges of our 200 \AA binning are plotted as grey lines, with the integrated values for those bins over plotted. } \label{spec_w4}
\end{figure}

\begin{deluxetable}{cccccc}  
\tabletypesize{\small}
\tablecolumns{4}
\tablecaption{WASP-4b: Calibrator Stars \label{w4cals}}
\tablewidth{0.4\textwidth}
\tablehead{
	\colhead{Identifier} 		& 
	\colhead{R.A.} 			& 
	\colhead{Dec.} 			& 
	\colhead{V mag.} 		&
	\colhead{R mag.} 		&
	\colhead{I mag.} 		}
\startdata
WASP -4   	&	23:34:15.09	&	-42:03:41.05			& 12.48	&	11.90	&	12.09	\\
Cal \#1		&	23:35:09.73     	&	-41:54:36.47 			& 13.24 	&	13.07	&	12.90	\\   
Cal \#2		&	23:35:34.59    	&	-41:52:34.75  			& 12.09 	&	11.86	&	11.52	\\ 
Cal \#3		&	23:35:41.22   	&	-41:57:40.27 			& 11.77 	&	11.41	&	11.01	\\  
Cal \#4		&	23:35:27.70	&	-42:04:15.13			& 13.07 	&	12.99	&	12.84	\\  
Cal \#5		&	23:35:44.41	&	-42:06:19.16			& 12.41 	&	12.40	&	12.40	\\	
Cal \#6		&	23:36:03.58   	&	-42:06:10.32 			& 11.66 	&	11.52	&	11.37	\\ 
Cal \#7		&	23:35:37.94	&	-42:10:46.97			& 13.18 	&	13.04	&	12.87	\\	 
Cal \#8		&	23:35:09.24	&	-42:15:10.52			& 13.75 	&	13.57	&	13.39	\\	
Cal \#9		&	23:34:44.80	&	-42:14:05.85			& 12.50	&	12.28	&	12.04	\\	
Cal \#10		& 	23:34:18.36	& 	-42:04:50.99			& 12.93	&	12.76	&	12.56	\\
Cal \#11	 	& 	23:34:04.52 	& 	-42:02:43.22 			& 14.52 	&	14.31	&	14.07				
\enddata
\tablecomments{See Figure \ref{FOVS} for a visual representation of the layout of the calibrator stars.}
\end{deluxetable}

\subsection{The planet WASP-52b} \label{W52_obs}
WASP-52b is a $\sim$1300K inflated Hot Saturn orbiting a K-type star. WASP-52b has been targeted for transmission spectroscopy by \cite{Kirk2016}, \cite{Mancini2017}, \cite{Chen2017}, and \cite{Louden2017}.  In addition, \cite{Tsiaras2017} recently analyzed archival Hubble transits of WASP-52b. \citeauthor{Kirk2016} uses wide band filters on the William Herschel Telescope, while \cite{Mancini2017} uses multiple medium-sized telescopes to obtain 13 light curves with wide band filters. In this work, we report higher spectral resolution light curves, with resolution similar to that reported in \cite{Chen2017} and \cite{Louden2017}. \cite{Chen2017} report a detection of the sodium line with a single transit observed from the Gran Telescopio Canarias from 5220 \AA{ }to 9030 \AA. \cite{Louden2017} similarly cover the entire optical range, from 4000 \AA{ }to 8750 \AA, with two discrepant transits observed from the William Herschel Telescope. The Hubble observations reported in \cite{Tsiaras2017} cover 1$\mu m$ to 1.6$\mu m$ and are binned at higher spectral resolutions than ground-based optical studies. Because of the spotty nature of WASP-52, additional transits of WASP-52b allow us to provide tighter constraints on the atmosphere's transmission spectra.
\par In this work, WASP-52b was observed on the nights of August 11th and September 21st, 2016 with exposure times of 70 seconds for both nights, resulting in an observational efficiency of 46\%. We obtained 154 exposures from the first night and 170 exposures on the second night. Seeing was very good on night 1, ranging from 0.6'' to 0.9''. The second night had high wind throughout, with seeing generally above 1.0'' but below 1.5''. The airmass of WASP-52 ranged from $\sim$2.5 $\rightarrow$ $\sim$1.2 $\rightarrow$ $\sim$1.7 on the first night, with similar values on the second night. As with WASP-4b, we are unconcerned about these seeing conditions due to our wide slits. In addition to our main target, our instrument mask allowed us to simultaneously target 8 calibrator stars (see Table \ref{w52cals}). One calibrator is reserved as a `check' star to ensure our pipeline accurately accounts for airmass and seeing effects throughout the night

\begin{figure}
	\epsscale{1.2}
	\plotone{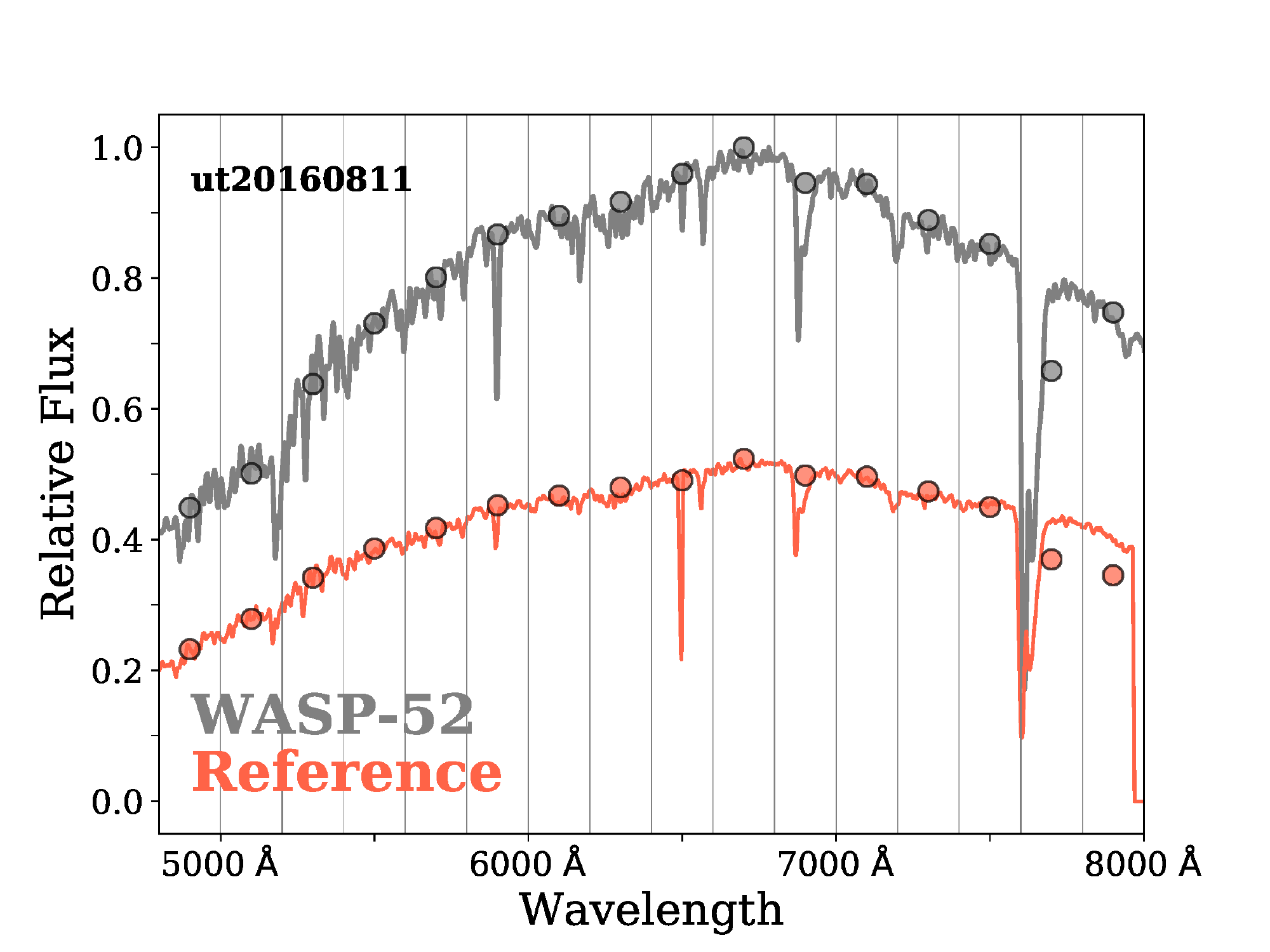}
	\plotone{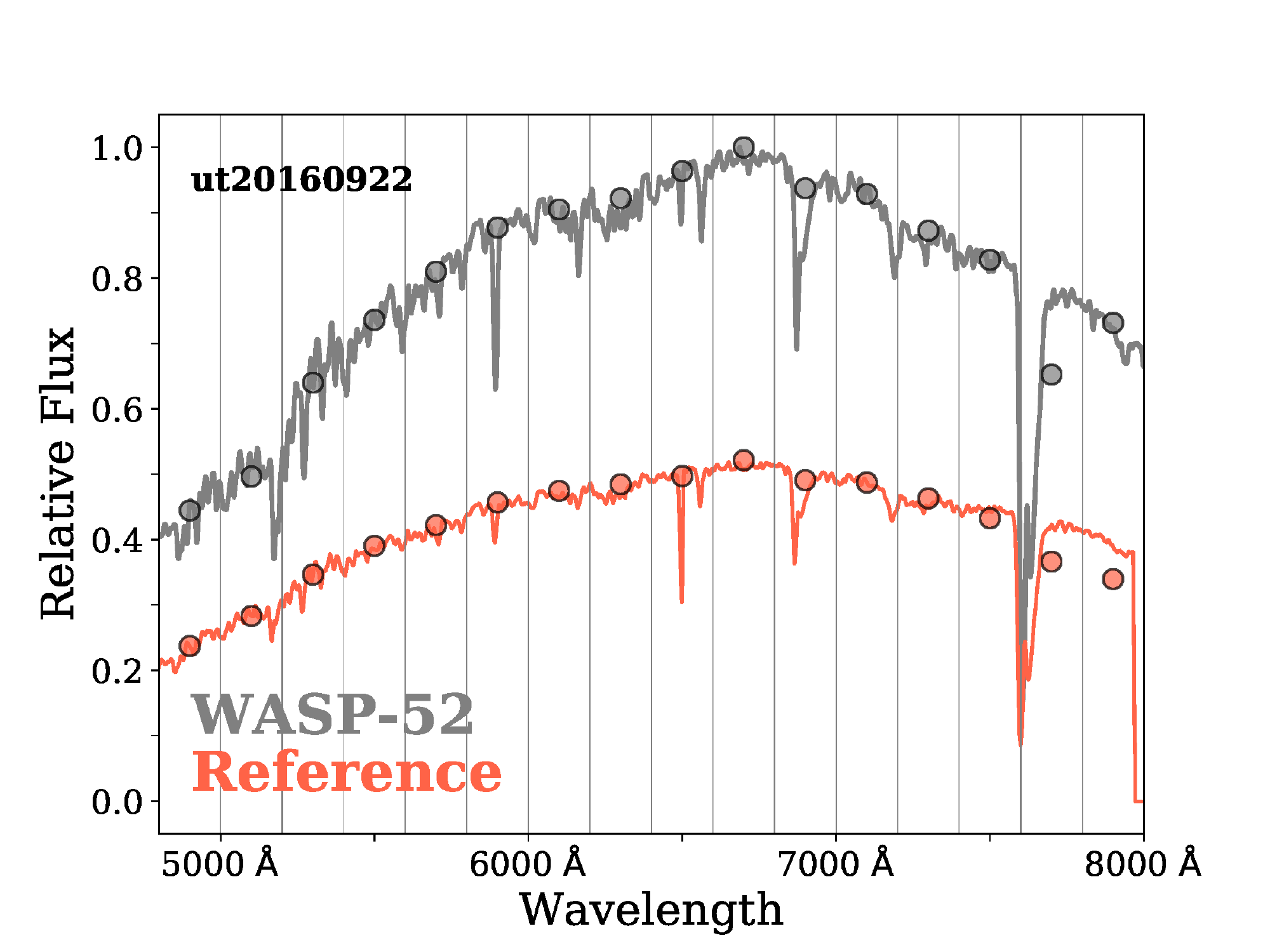}
	\caption{Spectra for each night of WASP-52 observations from 4600 \AA through 8200 \AA. The grey line shows WASP-52, and the red line shows the calibrator. The edges of our 200 \AA binning are plotted as grey lines, with the integrated values for those bins over plotted. } \label{spec_w52}
\end{figure}

\begin{deluxetable}{cccccc}  
\tabletypesize{\small}
\tablecolumns{4}
\tablecaption{WASP-52b: Calibrator Stars \label{w52cals}}
\tablewidth{0.4\textwidth}
\tablehead{
	\colhead{Identifier} 		& 
	\colhead{R.A.} 			& 
	\colhead{Dec.} 			& 
	\colhead{V mag.} 		&
	\colhead{R mag.} 		&
	\colhead{I mag.} 		}
\startdata
WASP-52		&	23:13:58.76	& 	+08:45:40.60		& 12.00	&	11.98	&	11.79	\\
Cal \#1		&	23:14:23.71     	&	+08:27:56.76 		& 12.21 	&	12.99	&	12.92	\\   
Cal \#2		&	23:14:16.72    	&	+08:42:16.92 		& 11.76 	&	12.73	&	12.47	\\ 
Cal \#3		&	23:14:07.26   	&	+08:39:29.28 		& 12.26	&	11.74	&	11.39	\\  
Cal \#4		&	23:13:31.96	&	+08:40:55.47		& 11.25 	&	12.03	&	11.88	\\  
Cal \#5		&	23:13:19.57	&	+08:30:35.68		& 11.22 	&	11.93	&	11.84	\\	
Cal \#6		&	23:13:29.15   	&	+08:45:10.21 		& 11.60 	&	12.33	&	12.22	\\ 
Cal \#7		&	23:13:17.19	&	+08:38:25.27		& 12.43 	&	12.89	&	12.68	\\	 
Cal \#8		&	23:13:20.30	&	+08:47:17.94		& 11.30 	&	12.01	&	11.89							
\enddata
\tablecomments{See Figure \ref{FOVS} for a visual representation of the layout of the calibrator stars.}
\end{deluxetable}


\section{Data Analysis} \label{analysis}

\subsection{Reduction Pipeline} \label{pipeline}
Our data reduction pipeline was developed in python from scratch in order to better understand the various sources of error. Our pipeline follows the standard spectroscopic reduction techniques as outlined in this section.
\par All data is first dark-corrected using dark frames taken in the afternoon prior to each observation. For each night, we took approximately one dozen dark frames, with exposure length equal to the science frames, which are combined into a median-filtered master dark frame before being subtracted. The variance in the master-dark is taken to be a source of error in pixel count levels. 
\par To begin data extraction, our pipeline uses flat frames taken with the mask and dispersing grism in place to identify the extent of the spectra on the chips for each object observed. A box is generated encompassing this entire region, which corresponds to approximately 200 pixels wide by nearly 3000 pixels long. This process accounts for curvature in the spectrum by extending the width of the extracted 2D region as needed if there is curvature detected. Because the size of the target is much smaller than the size of our slits, the width of these 2D boxes is always more than sufficient to capture the entire spectrum. For each target, the 2D regions are stitched across the y-direction chip gap of 57 pixels. In each 2D spectrum, we perform cosmic ray rejection following the process described in \cite{Nikolev2014}. We do not find that cosmic rays are a problem in our data, but use this step to ensure that we remove as many sources of error as possible
\par From the 2D spectrum, we first account for outliers in the background in order to improve our background fit. This is done row-by-row searching for `flared' pixels more than 3$\sigma$ the surrounding median background level, while masking the central region consisting of the object's spectrum. Outlier background pixels are mapped to the median of the surrounding pixels. A first-order polynomial is then fit to the background and subtracted off. Next, for each row in spectral space of the 2D spectrum, a gaussian is fit to the data to determine the x-centroid and size of the aperture (from the FWHM) to be used for each exposure. We again map 3$\sigma$ outliers from the gaussian fit to the surrounding median and then re-fit the gaussian to ensure there are no outliers significantly affecting the x-centroid and FWHM. It is important to note is that we do allow the aperture size to vary from exposure to exposure, but it is constant across all wavelengths for a given exposure (We find this method useful to account for variations in seeing and airmass throughout the night). Each exposure's aperture is computed from the median FWHM across all spectral rows for each individual exposure and object. The aperture size is defined to be 3$\times$ the median FWHM for each exposure and object. Shifts in the star's position in the spatial direction on the chip are accounted for by allowing the central position of the gaussian to float for each row, but in reality, the centroid also remains fairly constant throughout the extraction, due to the stability of the pointing of the Baade telescope. Finally, we sum the region within the aperture size around the x-centroid, accounting for the 57 chip gap by assigning non-number values to those rows in spectral space (This later allows us to discard the binned spectra including this region). 
\par Prior to applying a wavelength solution, we first account for shifts in the star's position in the spectral direction on the chip. This involves oversampling each spectrum in the time-series by a factor of 10 and correlating them with the first exposure to measure pixel shifts in the spectral direction down to 1/10$^{\mathrm{th}}$ of a pixel. Just like we previously account for stellar shifts in the spacial direction on this chip by fitting the x-centroid location, this step allows us to correct for stellar shifts in the spectral direction. Each exposure is then shifted to the same pixel space as the first exposure as needed. 
\par Because the large size of our target slits would result in our arc frames being saturated at the shortest exposure time, here we instead use small 1'' `mini-slits' directly to the side of our object slits for wavelength calibration. The IMACS instrument has available a HeNeAr lamp with 24 identified lines from $\sim$ 3900\AA{ } to $\sim$8500\AA. Our pipeline requires an initial manual marking of several lines and from there extrapolates a polynomial with a user-defined order while matching up additional lines. The user can then accept or reject each newly matched line. Because a line matched by the pipeline must be within 0.25 pixels of the known arc spectrum, obvious matches are sometimes not identified by the routine. At this point, the user can manually feed in the locations of these additional lines to improve the fit. Once the user accepts a fit, a pixel-to-wavelength conversion is calculated and applied to the spectra for that object. This process is done for each object on the mask (the main target and each calibration star). Because the lines below $\sim$5000\AA are weak in our arcs, we do not currently use the wavelength solution below approximately $\sim$4600\AA. 
\par We finally bin the spectra for each star into 200\AA{ }wide bins from 4600 - 8200\AA, for a total of 18 bins; 100\AA wide bins, for a total of 36 bins; as well as 20\AA{ }wide bins, for a total of 180 bins. For this step, we make use of the python package SpectRes \citep{Carnall2017}. Bins containing a chip gap and/or the large O$_2$ absorption near 7600\AA{ }are discarded. The wide binning allows for measurements of any Rayleigh slope, while minimizing noise in the light curves. The finest binning allows for detection of Alkali absorption features. Applying a variety of binning ensures we return the overall shape of the transmission spectrum, while not averaging over any narrower features. Figures \ref{spec_w4} and \ref{spec_w52} show spectra from one point in time for all three nights at the widest binning, as well as the combined calibrator star. 

\subsection{Light Curves}
\par
\begin{figure}
	\epsscale{1.2}
	\plotone{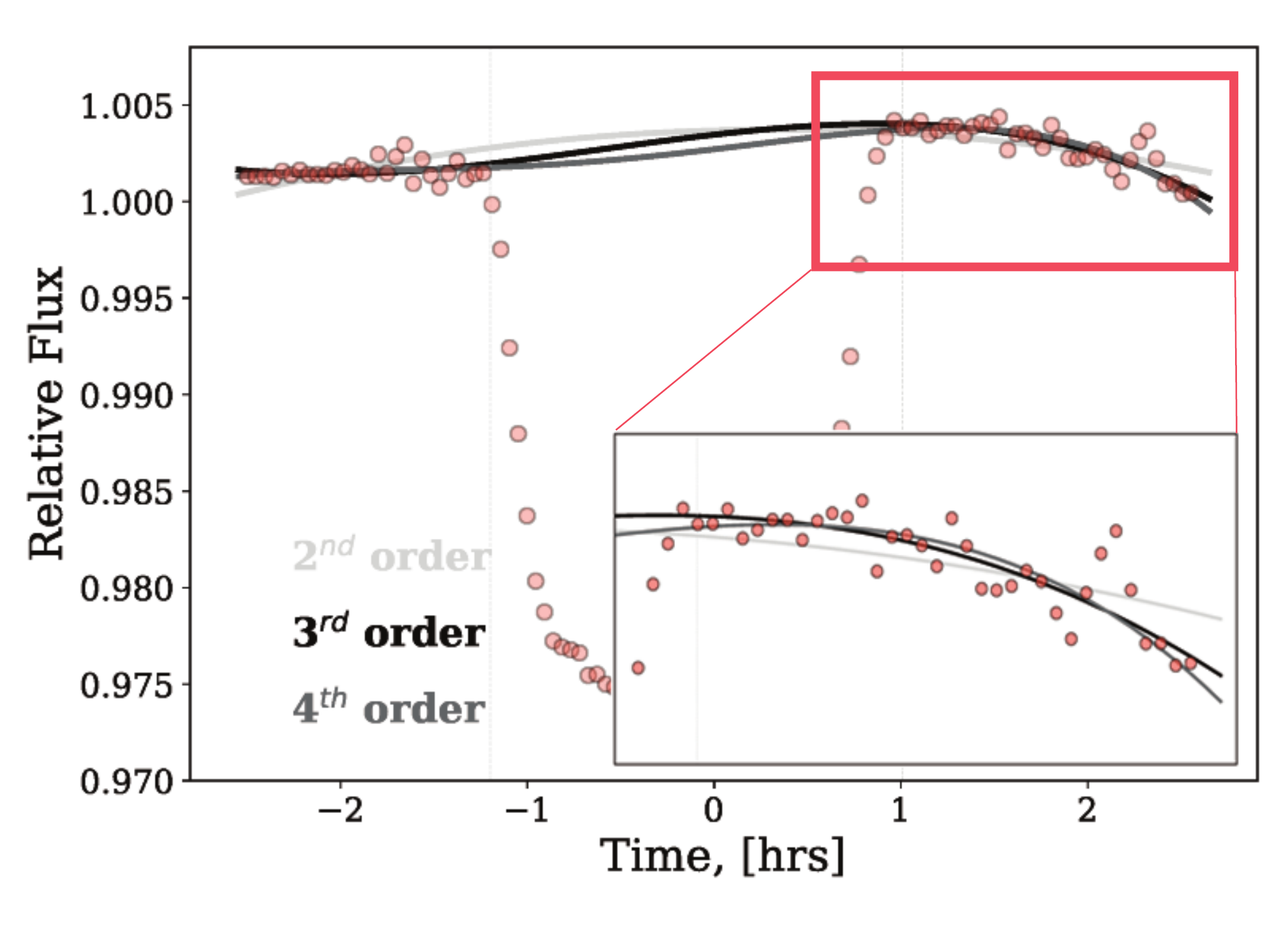}
	\caption{White light curve for WASP-4b, overlaid with several baseline models, all polynomials in time of varying order. The inset shows a zoomed-in portion of the light curve after transit, demonstrating that the 3$^{rd}$ and 4$^{th}$ order fits are similar, but that the 2$^{nd}$ order does not properly capture the baseline. While the 3$^{rd}$ order and 4$^{th}$ order fits are similar, we choose the 3$^{rd}$ order fit to minimize free parameters.} \label{baseline}
\end{figure}

 To generate our binned and white light curves, we first create `master' calibrators. Each master calibrator consists of all reference stars, minus the one chosen as the `check' star. This master calibrator is divided out for both our target star and the check star before an out-of-transit time series model is fit. We fit both models of time (2$^{nd}$, 3$^{rd}$, and 4$^{th}$ order polynomials) and airmass (3$^{rd}$ order polynomial). We determine the 3$^{rd}$ order polynomial in time to be the best reproduction of our observed baseline after dividing out our calibrator stars. Figure \ref{baseline} shows our white light curve for WASP-4b, with several baseline models in time. Although the 3$^{rd}$ and 4$^{th}$ order polynomials show similar fits, we choose the 3$^{rd}$ order fit to minimize the number of free parameters involved when fitting the binned light curves which naturally are noisier.
\subsubsection{White Light Curves}
\par White light curves are generated from the entire spectral range, covering 4600\AA{ } to 8200\AA. These white light curves are used to fit for orbital parameters using the python package Batman\citep{Kreidberg2015} with a quadratic limb-darkening function, in conjunction with a MCMC fitting routine, emcee \citep{FMackey2013}. We adopt the quadratic limb darkening coefficients from \cite{Claret2011} as starting points for the fits based on the stellar parameters in Tables \ref{W4-params} and \ref{W52-params}. We fit for the center of transit, period/semi-major axis, inclination of orbit, white-light planet radius, and white-light limb darkening, where orbital and stellar parameters are fit with a gaussian prior defined by the values and errors given in Tables \ref{W4-params} and \ref{W52-params}. The relation between period and semi-major axis (in units of stellar radii) is maintained for the assumed stellar mass and radii values by only fitting for period, while simultaneously calculating a corresponding semi-major axis with a stellar mass and radii drawn from a gaussian distribution around the errors defined by the parameters in Tables \ref{W4-params} and \ref{W52-params}. 
\par
\begin{figure}
	\epsscale{1.1}
	\plotone{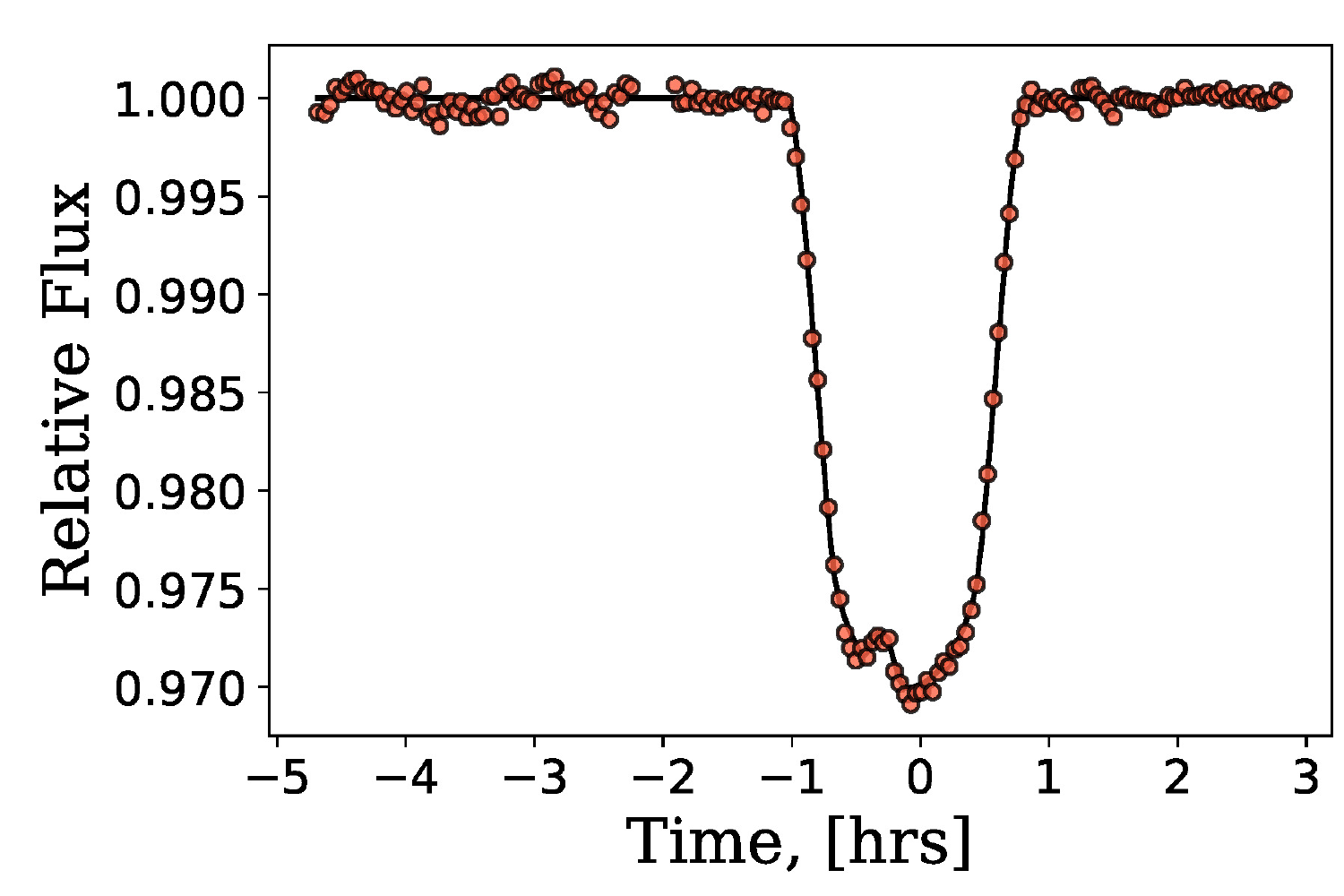} \\ 
	\caption{White light curve for the second transit of WASP-52b, showing a clear spot crossing feature at approximately -0.5 hours.} \label{w52_2_wlc}
\end{figure}

 For the second transit of WASP-52b, there is a clear star spot crossing event visible in the white light curve (see Figure \ref{w52_2_wlc}). In exoplanet transits light curves, star spots appear as `bumps' in transit due to the planet  blocking out less light while occulting a dark spot. Spots occulted during transit require fitting a light curve with 4 additional parameters per spot; spot x and y location relative to the stellar center, spot radius, and spot contrast. Similarly to orbital parameters, the spot location and size will not change as a function of wavelength. We employ a python package named SPOTROD \citep{Beky2014}, which generates light curve models in combination with any number of spot occultations. Although our first transit of WASP-52b does not show a star spot feature, we chose to model both transits with this package for consistency. For both transits, the white light curve fitting searches for the best orbital parameters and star spot location and size parameters, with the expectation that the spot radius should be 0 for the first transit.
\par In order to ensure that the joint spot-orbital parameter fitting does not alter our orbital parameters, we also run white light curve fits for just orbital parameters with the spot masked in the data. We find that both techniques provide the same orbital parameters, within errors. For all transits, the orbital parameters are then assumed to be constant with wavelength. For spotted transits, spot location and size are held constant across wavelengths as well.
\par Additionally, we preform the same analysis on our check stars from each night. In these cases, the light curves should be best fit by a flat line, finding no planet transit. In all cases, this is found to be true.
\subsubsection{Binned Light Curves}
\par
\begin{figure*}
	\epsscale{1.1}
	\plotone{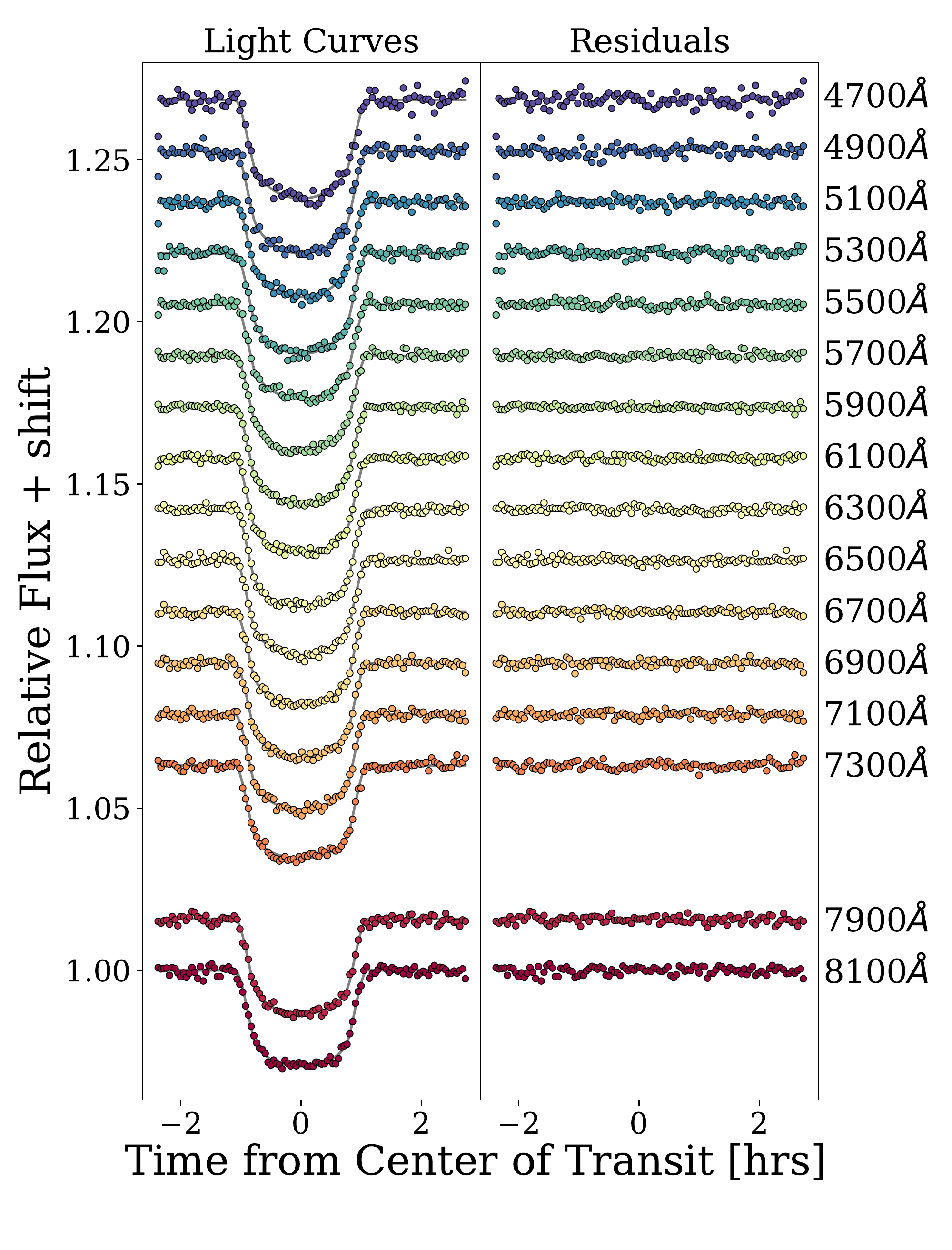} \\ 
	\caption{Light curves for the low resolution binning (200\AA) for WASP-4b. The left frame shows our reduced light curves and best fit model in black. The right panel shows the residuals for the fits. We do not include bins that are contaminated by major O$_2$ absorption in Earth's atmosphere.} \label{lcs_w4}
\end{figure*}

\begin{figure*}
	\epsscale{1.1}
	\plotone{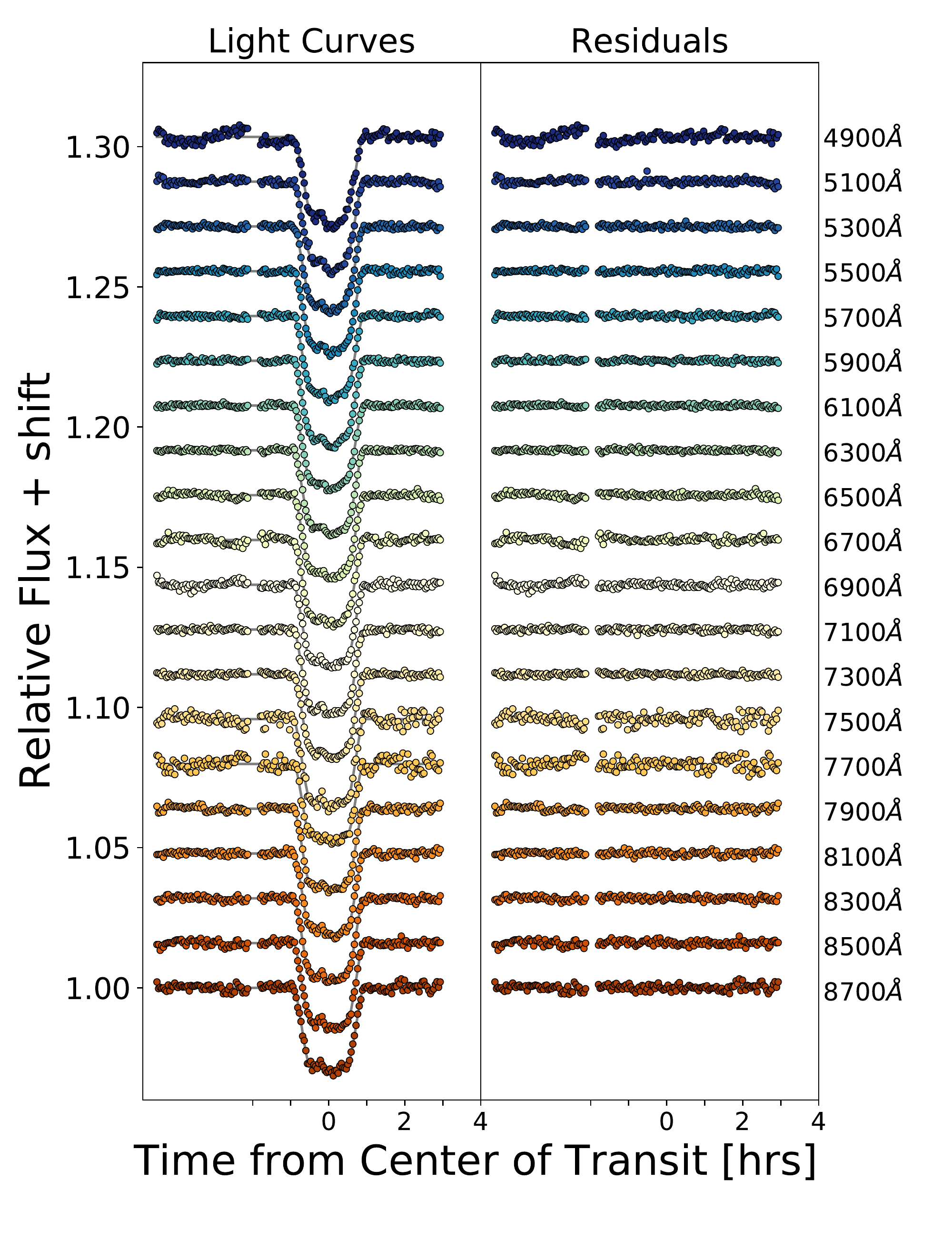}
	\caption{Light curves for the low resolution binning (200\AA). Top: WASP-52b Transit 1 (unspotted, 08/11/2016). Bottom: WASP-52b Transit 2 (spotted, 09/22/2016). For both transits, the left frame shows our reduced light curves and best fit model in black. The right panel shows the residuals for the fits. We do not include bins that are contaminated by major O$_2$ absorption in Earth's atmosphere.} \label{lcs_w52}
\end{figure*}

\begin{figure}
	\epsscale{1.1}
	\plotone{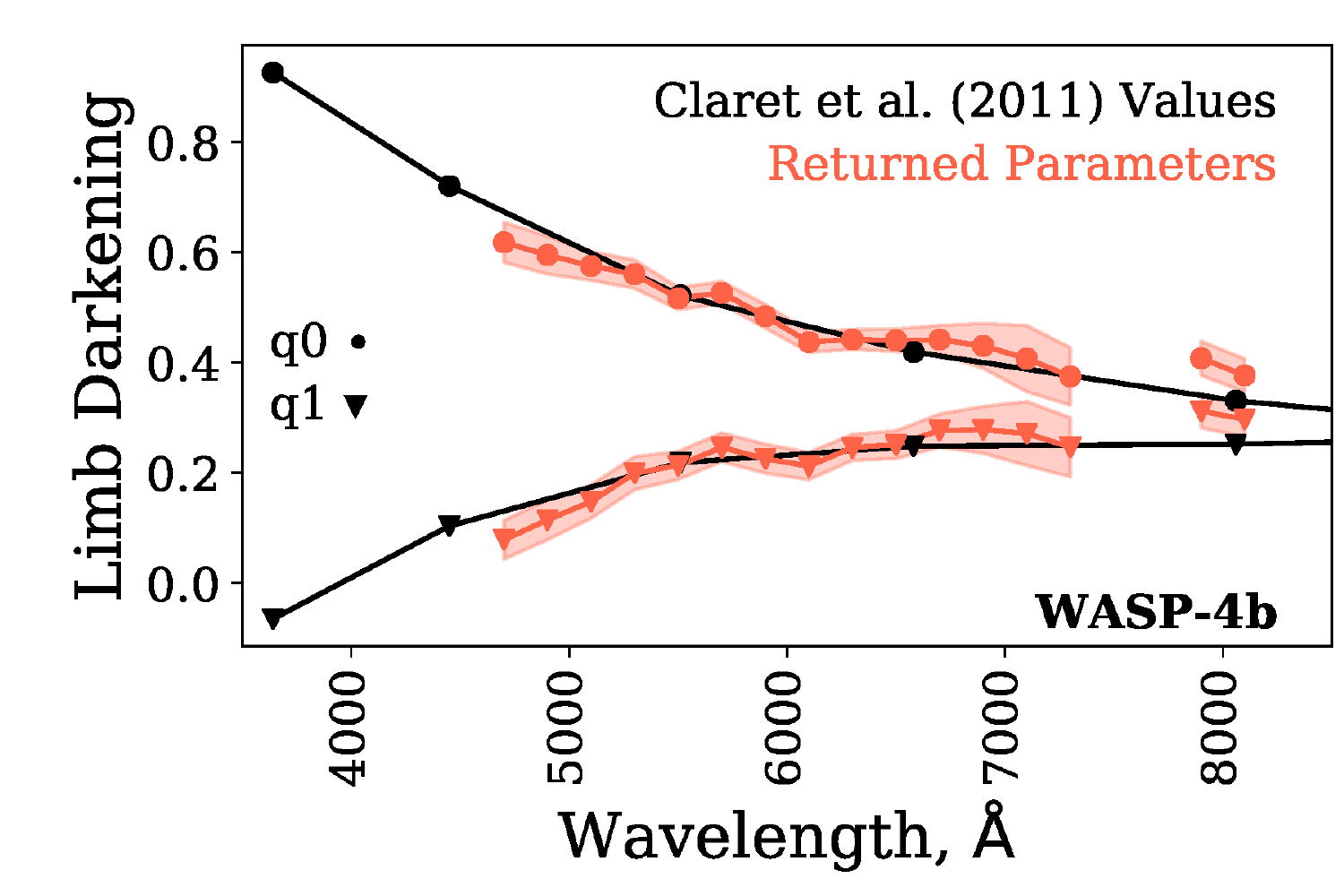}\\
	\vspace{0.2cm}
	\plotone{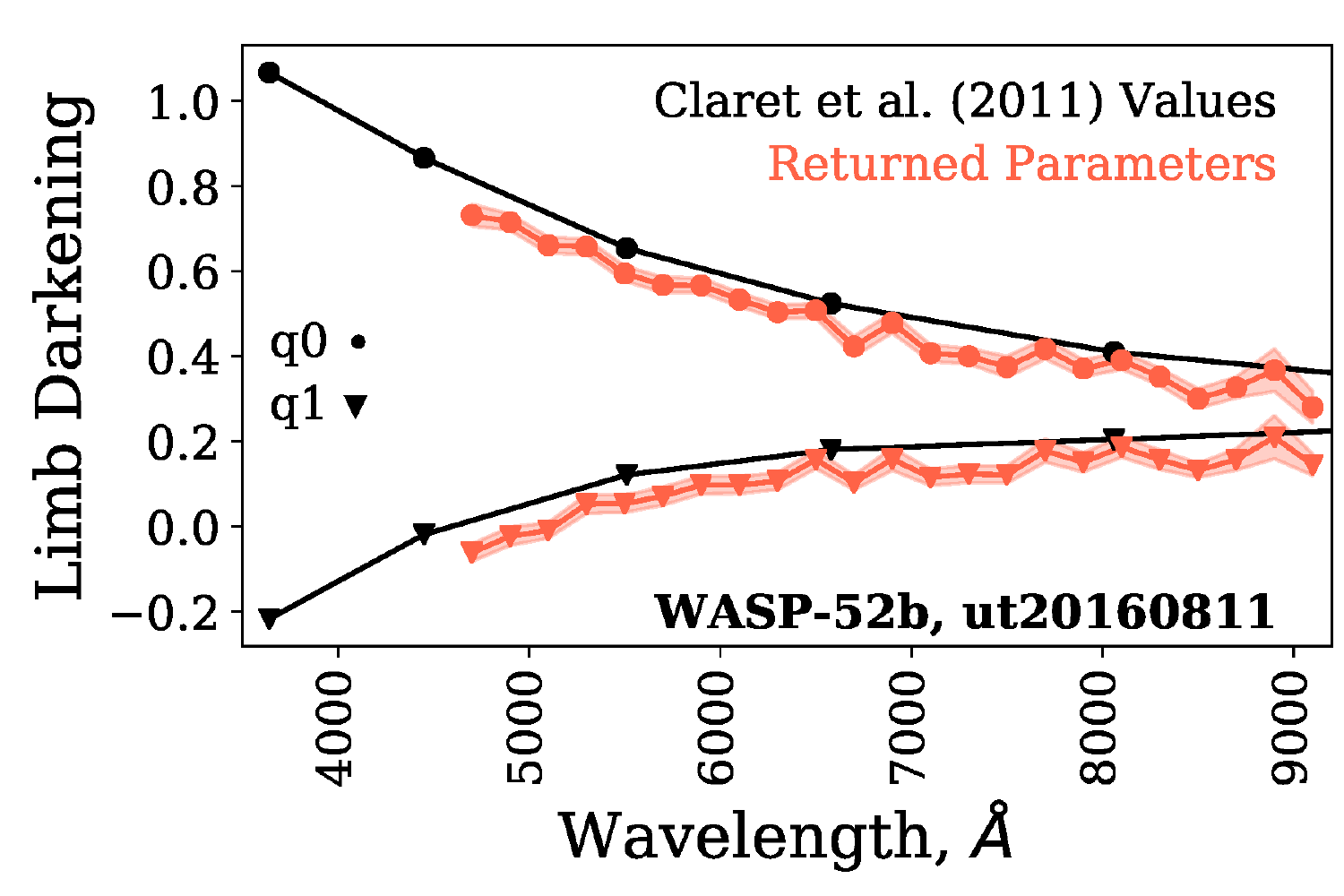} \\ 
	\vspace{0.2cm}
	\plotone{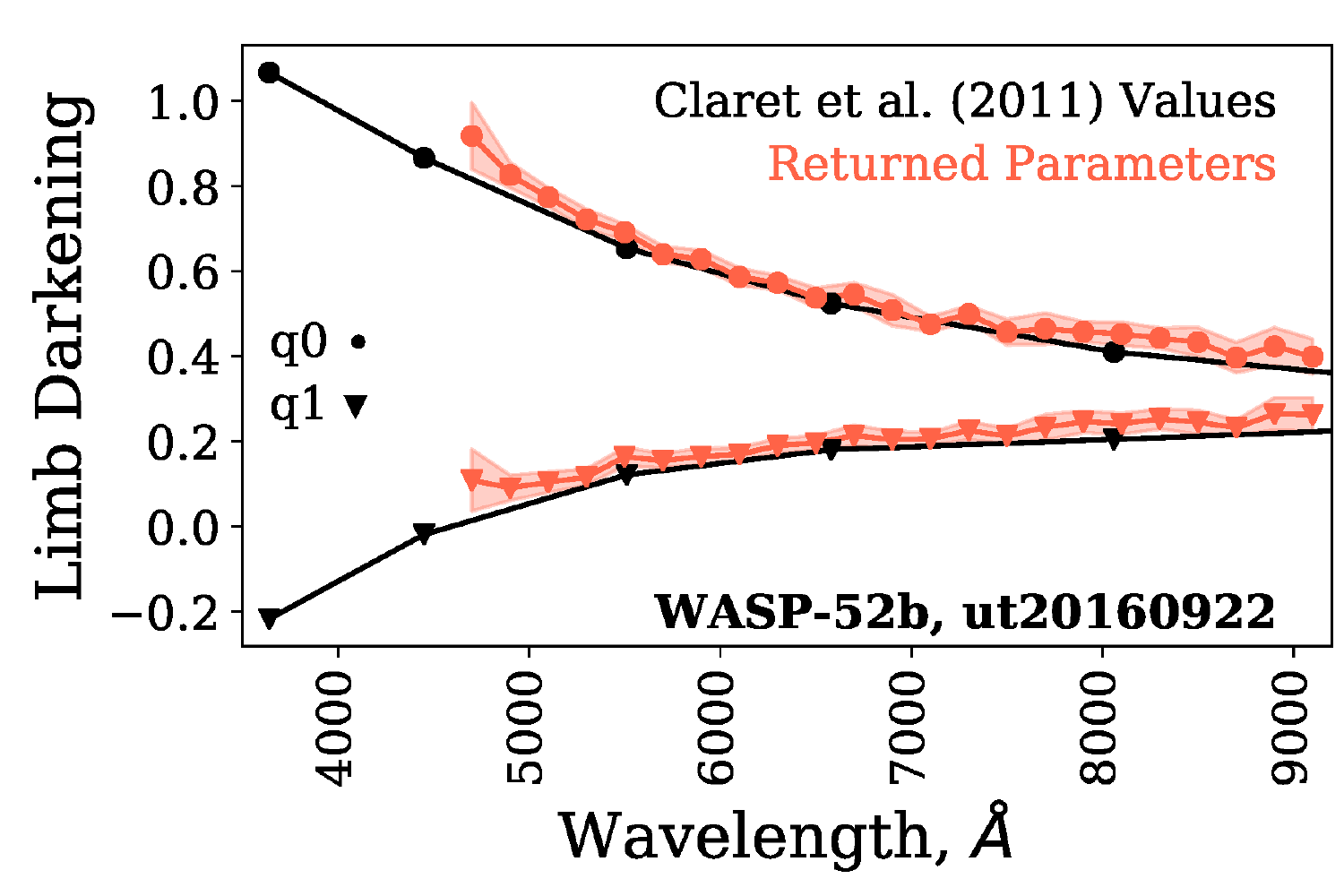}
	\caption{Limb darkening parameters from MCMC fits (red points) compared to literature values (black curves). All three transit events are shown.} 
	\label{claret_ld}
\end{figure}

 As discussed in Section \ref{pipeline}, we bin the data into 200\AA, 100\AA, 50\AA, and 20\AA{ }wide bins in wavelength space. Light curves and their fits are shown in Figures \ref{lcs_w4} and \ref{lcs_w52}. For each binning we perform MCMC fits using either Batman (WASP-4b, no stellar spots) or SPOTROD (WASP-52b, spot crossing events), as well as noise removal as discussed in \ref{noise}. Because orbital parameters are fit using only the white light curve, we need only fit for radius and limb darkening coefficients for WASP-4b in the binned light curves, and the same plus spot contrast for the binned WASP-52b light curves. For each bin, the quadratic limb darkening coefficients are interpolated between the Johnson filter values from \cite{Claret2011} and fit together across bins by requiring that the shape of the limb darkening wavelength-dependent interpolation remains approximately constant, while allowed to shift to better fit the data and to account for uncertainties in the stellar parameters and limb darkening tables. For WASP-4b and the first transit of WASP-52b, our fit limb darkening values agree with those from \cite{Claret2011} For the second transit of WASP-52b, see Section \ref{limbspots}. Figure \ref{claret_ld} shows our fit limb darkening values as compared to the \cite{Claret2011} values, which demonstrates our fitting procedure, where, for all three transits, the limb darkening parameters show the same wavelength-dependent shape as the \cite{Claret2011} values, but may be systematically shifted higher or lower to account for the uncertainties in the stellar parameters and limb darkening tables.
\begin{figure*}
	\epsscale{1.1}
	\plotone{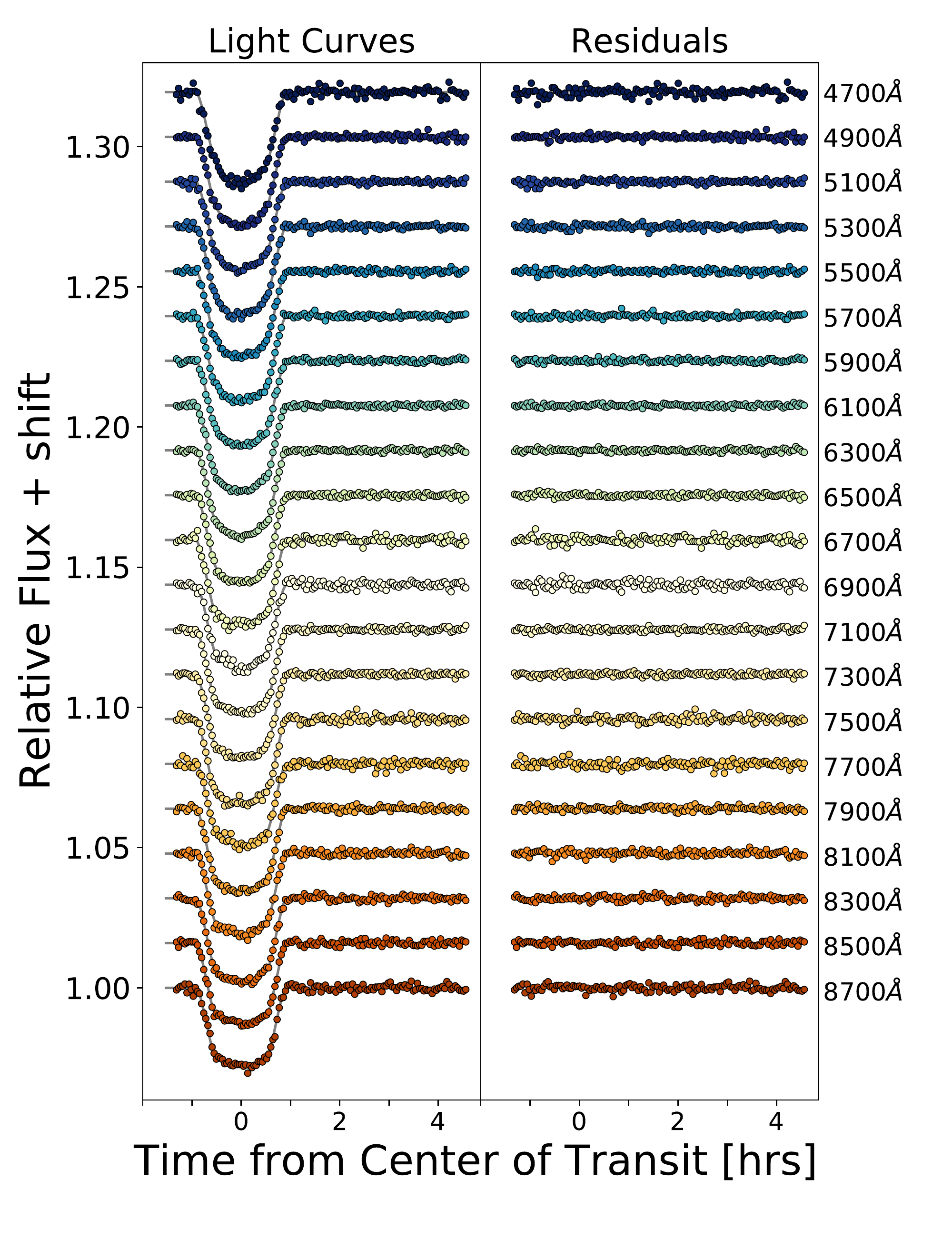} \\ 
\end{figure*}
\begin{deluxetable}{cc} 
\tabletypesize{\small}
\tablecolumns{5}
\tablecaption{WASP-4b: Stellar and Orbital Parameters \label{W4-params}}
\tablewidth{0.8\textwidth}
\tablehead{
	\colhead{Stellar Parameter}		&
	\colhead{Value}			           	}
\startdata
Mass, [M$_{\odot}$]			&	0.89$\pm$0.01		\\
Radius, [R$_{\odot}$]		&	0.92$\pm$0.06  	\\
T$_{eff}$, [K]				&	5436$\pm$34		\\
{ }[Fe/H], [dex]				&	-0.05$\pm$0.04		\\
$\xi$, [km/s]				&	0.85$\pm$0.10$^{\dag}$	\\
{\scriptsize{(Microturbulent Velocity)}} & { }				\\
$\log{g}$, [cm/s]			&	4.28$\pm$0.06		\\ \hline \hline
Planet Parameter			&	Value	\\ \hline
Mass, [M$_{Jup}$] 	& 1.22$\pm$0.01				\\
Radius, [R$_{Jup}$] 	& 1.33$\pm$0.16				\\
T$_{eq}$, [K]			& 1664$\pm$54				\\
Period, [days]		& 1.3382325$\pm$3e-7				\\
Semi-major axis, [AU]	& 0.0232$\pm$0.0005$^{\star}$		\\
Eccentricity			& 0.0							\\
Inclination, [deg]		& 86.85$\pm$1.76			\\ 				
\enddata
\tablecomments{All values have been adopted from \cite{Petrucci2013} with the exception of those marked with a $^{\star}$, which come from \cite{Southworth2012}, and those marked with a $^{\dag}$, which come from \cite{Doyle2013}.}
\end{deluxetable}
\begin{deluxetable}{cc} 
\tabletypesize{\small}
\tablecolumns{5}
\tablecaption{WASP-52b: Stellar and Orbital Parameters \label{W52-params}}
\tablewidth{0.8\textwidth}
\tablehead{
	\colhead{Stellar Parameter}		&
	\colhead{Value}			           	}
\startdata
Mass, [M$_{\odot}$]			&	0.87$\pm$0.03		\\
Radius, [R$_{\odot}$]		&	0.79$\pm$0.02  		\\
T$_{eff}$, [K]				&	5000$\pm$34			\\
{ }[Fe/H], [dex]				&	0.03$\pm$0.12			\\
$\xi$, [km/s]				&	0.9$\pm$0.1			\\
{\scriptsize{(Microturbulent Velocity)}} & { }					\\
$\log{g}$, [cm/s]			&	4.58$\pm$0.01			\\ \hline \hline
Planet Parameter 			&  	Value				\\ \hline
Mass, [M$_{Jup}$] 		& 0.46$\pm$0.02				\\
Radius, [R$_{Jup}$] 	& 1.27$\pm$0.03				\\
T$_{eq}$, [K]				& 1315$\pm$35				\\
Orbital Period, [days]			& 1.749780$\pm$1e-6				\\
Semi-major axis, [AU]		& 0.0272$\pm$0.0003		\\
Eccentricity			& 0.0							\\
 Inclination, [deg]		& 85.35$\pm$0.20					
\enddata
\tablecomments{All values have been adopted from \cite{Hebrard2013}. } 
\end{deluxetable}
\subsection{Error Analysis and Noise Removal} \label{noise}
\par Our MCMC likelihood function takes the RMSE of the out-of-transit baseline as the input uncertainty on an individual point. From this, the error on our fit transit depths is the 1-sigma level from the converged MCMC chain. To further improve our results, we perform a noise removal algorithm to remove common mode white noise. We do not find a significant improvement when applying red noise removal techniques to these data sets. Our common mode white noise is the ratio between the baseline-uncorrected data and the white transit model fit, which we then divide out of our wavelength-binned light curves.
\section{Results} \label{results}

\subsection{Transmission Spectra}
\par For all transits, we produce transmission spectra based on the binned light curve fits. Our finer binning (100\AA, 50\AA, and 20\AA) serves to ensure that we do not miss absorption features that may be smeared out by wider binning (200\AA), however, these bins naturally result in larger error bars. We find no additional features at this binning and do not show light curves at this binning, though the points for the 100\AA{ }binning are shown on the transmission spectra in Figure \ref{W4_Tspec}. At optical wavelengths, we are primarily searching for signs of Rayleigh Scattering, as well as sodium and potassium absorption features. Rayleigh Scattering allows us to constrain the relation between the mean molecular mass of the atmosphere and the mean atmospheric limb temperature. The slope of this scattering is given by
\begin{equation}
\frac{1}{R_s}\frac{dR_p}{d\ln{\lambda}}=\alpha\frac{1}{Rs}\frac{k_BT}{\mu g}
\end{equation}
, first described by \cite{Lecavelier2008} where R$_s$ is the stellar radius; R$_p$ is the planet radius; $\alpha$ is the power of the wavelength dependance of Rayleigh scattering, typically -4; k$_B$ is the Boltzmann constant; T is the planetary limb temperature; $\mu$ is the mean molecular weight of the atmosphere; and g is the planetary gravity. Where, as noted, $\mu$ and T are our main unknown parameters. Even for planets with temperatures determined through secondary eclipse (such as WASP-4b), transmission spectroscopy probes the average temperature of the atmospheric limb, whereas secondary eclipse probes the average dayside temperature. While secondary eclipse measurements, in combination with expected equilibrium temperatures for an assumed albedo, can help break the degeneracy between the parameters, this degeneracy points to the necessity of multi-wavelength observations to fully understand the planetary atmosphere. It is important to note that large uncertainties in temperature observations still dominate the relation between the limb temperature and the mean molecular mass, and any statement about mean molecular weight must make an assumption for the temperature and albedo of the planet.
\par The sodium (Na) and potassium (K) absorption features occur at 5890\AA{ }and 7665\AA, respectively. For most planets, these are the only two absorption features detectable in the optical regime. It is important to note, however, that the detectability of these features heavily depends on the width of the wavelength bins chosen, as demonstrated by \cite{Chen2017}. Specifically calculated for WASP-52b, \cite{Chen2017} show that for bin sizes greater than $\sim$30\AA, no detectable increase due to these absorption features will be measured. Unfortunately, at these bin widths, the error in the measurements increases substantially. For this reason, we bin at 200\AA{ }to determine the overall shape of the transmission spectra, as well as at 100\AA{ } and 20\AA{ }to search for additional spectral features. We find no significant deviation from the 200\AA{ }binning in the higher spectral resolution light curves. 
\par Although the K absorption feature may be as detectable as Na, there is an oxygen absorption line in our atmosphere centered at $\sim$7600\AA. If bin sizes are wide enough that they contain the potassium absorption line and any part of this wide oxygen feature, the K feature may be diluted due to high noise.
\par Because our data is not of high enough resolution to perform atmospheric retrievals, we instead create representative transmission models using the code Exo-Transmit as described in \cite{Kempton2017}. Exo-Transmit is a robust and user-friendly package which allows the user to define the planetary parameters, as well as set the composition and temperature structure of the atmosphere. The user can set the pressure level of any cloud deck, and vary the strength of scattering in the atmosphere. We explore models with varying composition, scattering strength, and cloud decks. Throughout this work, we assume isothermal temperature structures.
\subsection{WASP-4b} \label{w4_results}
\begin{figure}
	\epsscale{1.3}
	\plotone{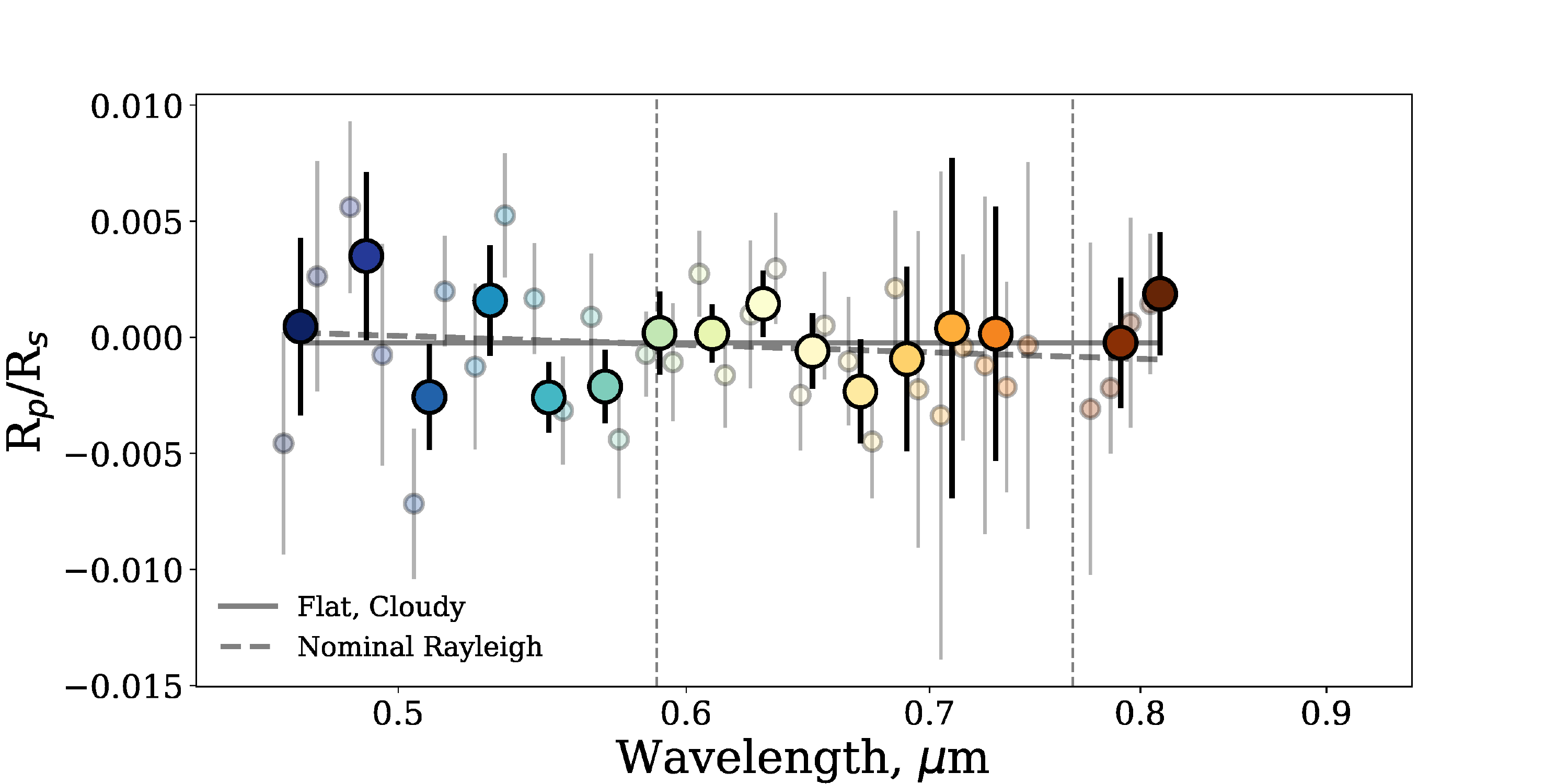} 
	\vspace{0.1cm}
	\plotone{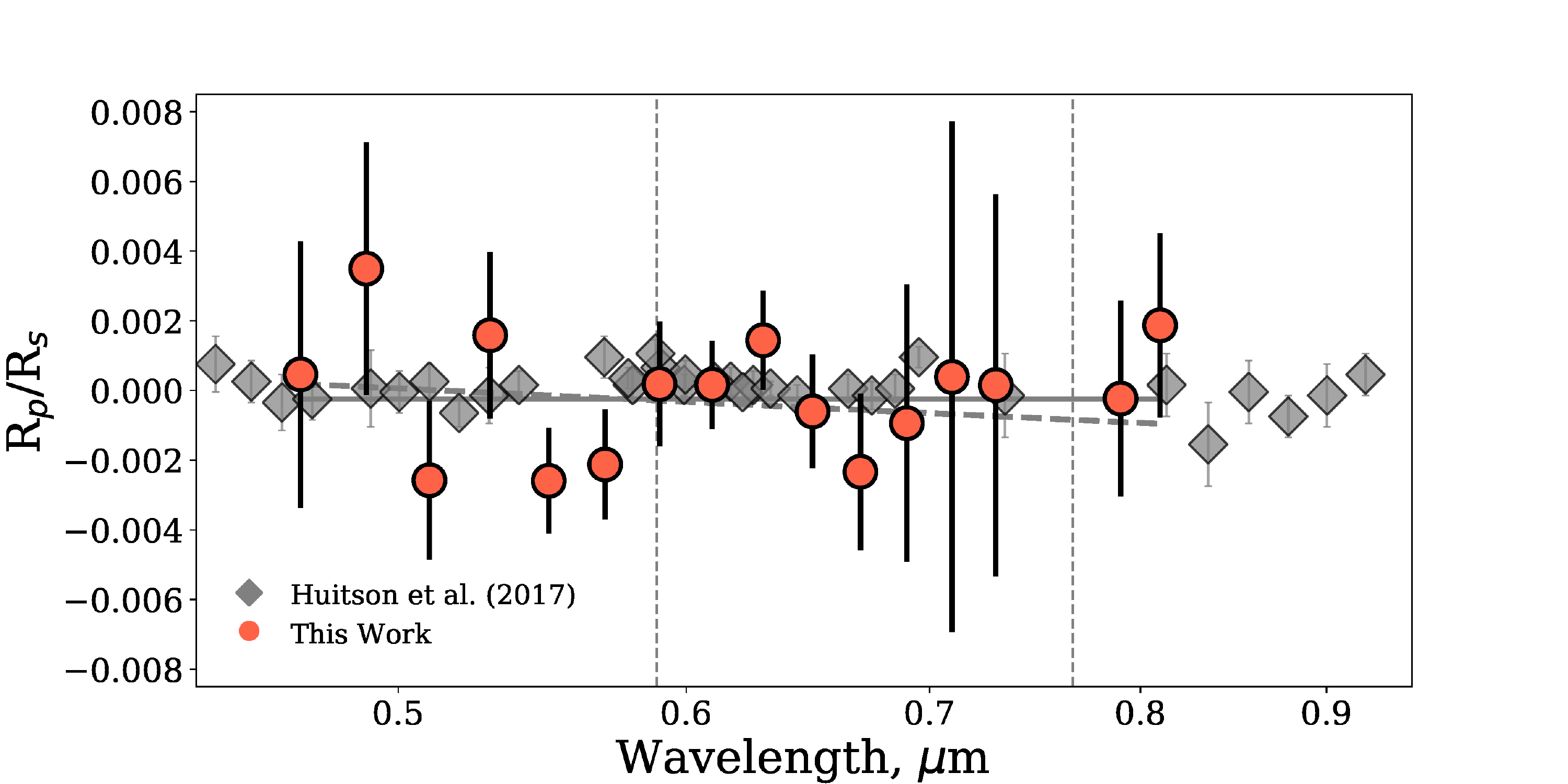}
	\caption{Transmission spectra for our one transit of the Hot Jupiter WASP-4b. For ease of comparison with values from \cite{Huitson2017}, values are plotted relative to the white light transit depth. \textbf{Top:} We present a Rayleigh slope at the equilibrium temperature of the planet (dashed line), and a flat spectra to mimic a cloudy atmosphere (solid line). We find that the models show nearly identical goodness-of-fit $\chi^{2}$ values, pointing to the necessity of additional longer wavelength data to distinguish between atmospheric models for such high gravity planets. \textbf{Bottom:} Our data in comparison to \cite{Huitson2017}, which agree well. We note that they obtained 3 transits of WASP 4-b, compared to our single transit.} \label{W4_Tspec}
\end{figure}

Figure \ref{W4_Tspec} shows our final transmission spectrum for WASP-4b. We note that within errors, WASP-4b presents a fairly flat, and featureless, spectrum; however one that is also consistent with Rayleigh scattering at the equilibrium temperature of the planet (assuming zero-albedo).  Observations into the infrared would help to clear up the degeneracy between models, and additional transits would improve the accuracy of measurements. Unfortunately, Hubble transit observations of this planet by \cite{Ranjan2014} were saturated and transit depth values were not reported. Additionally, WASP-4b's high gravity means transmission signals will be small, and so we do not expect single transit ground based observations alone to disentangle the atmospheric models. 
\par The cloudy model and the nominal Rayleigh slopes both return a nearly identical $\chi^2$ value, at $\sim$8$\times10^{-4}$. Based on our Exo-Transmit modeling of a solar metallicity atmosphere, we expect a cloud deck no lower than 10$^{-4}$ bar in order to obscure the Na and K absorption lines. 
\subsection{WASP-52b}
\begin{figure}
	\epsscale{1.3}
	\plotone{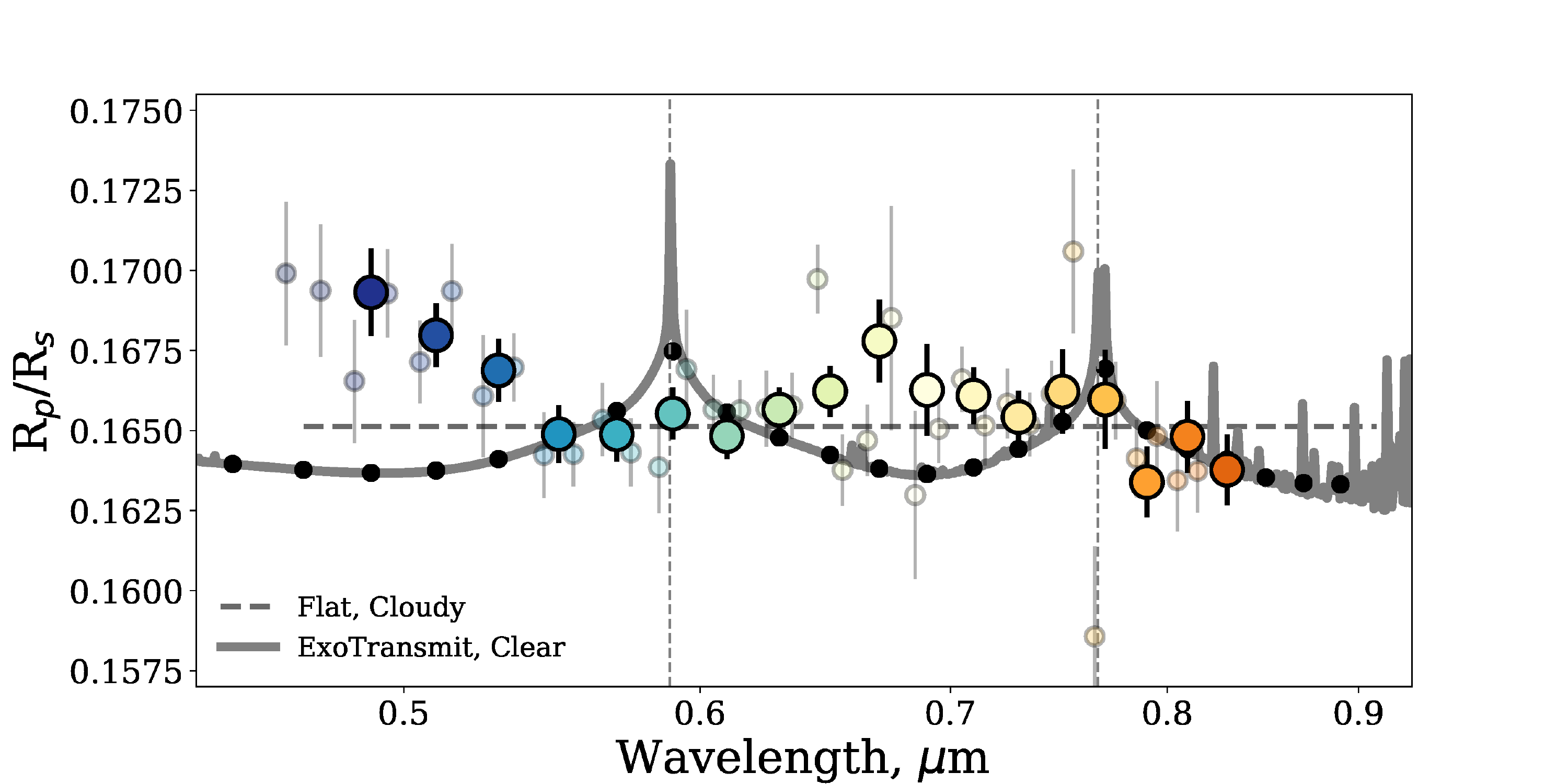} 
	\vspace{0.1cm}
	\plotone{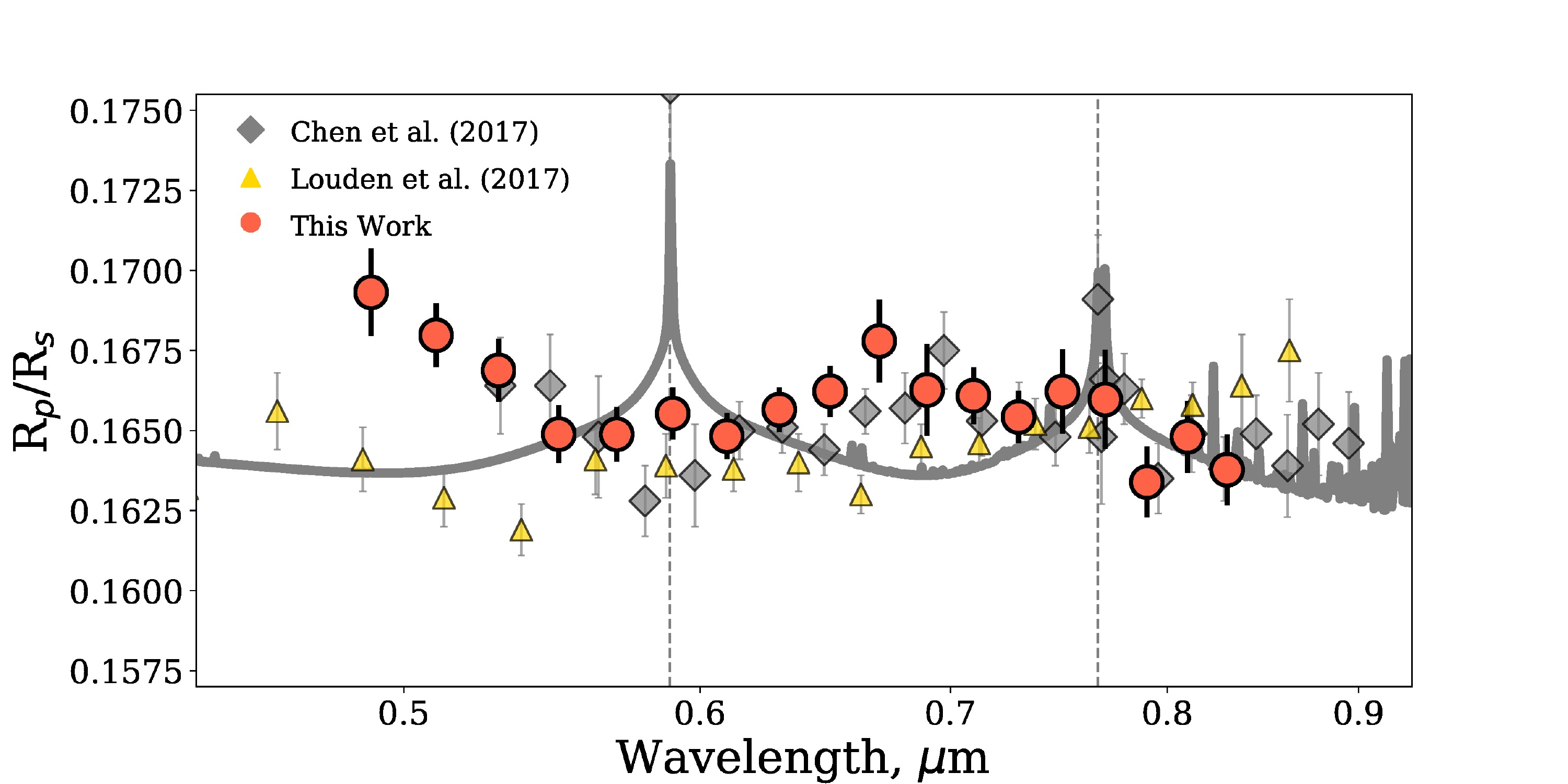}
	\caption{Combined transmission spectra for both transits of the Hot Saturn WASP-52b. \textbf{Top:} We present a linear fit to the data (dotted line) and an Exo-Transmit model with solar metallicity. We prefer a flat, cloudy spectra. \textbf{Bottom:} Our data in comparison to \cite{Chen2017} and \cite{Louden2017}, which agree well. } \label{W52_Tspec}
\end{figure}

Figure \ref{W52_Tspec} shows our combined transmission spectrum for WASP-52b, plotted with a linear fit and an Exo-Transmit model (solar metalicity, standard scattering). We find a large discrepancy in the average transit depth between the spotted transit (Night 2) and the unspotted transit (Night 1). The offset between a spotted and unspotted transit is also seen by \cite{Louden2017}. This difference may be a result of unocculted star spots on the stellar surface. See section \ref{unocspots} for a discussion of how one can correct for unocculted star spots. Further, we measure different limb darkening parameters on each night. Section \ref{limbspots} discusses one possible explanation for this difference. We do not find evidence of sodium absorption, as found by \cite{Chen2017}. Our data exhibits a flat transmission spectra, pointing to a cloudy atmosphere. Using Exo-Transmit models, we conclude that the cloud deck must be no lower than 10$^{-4}$ bar.

%
%
\subsubsection{Unocculted Star Spots}
\label{unocspots}
\begin{figure}
	\epsscale{1.0}
	\plotone{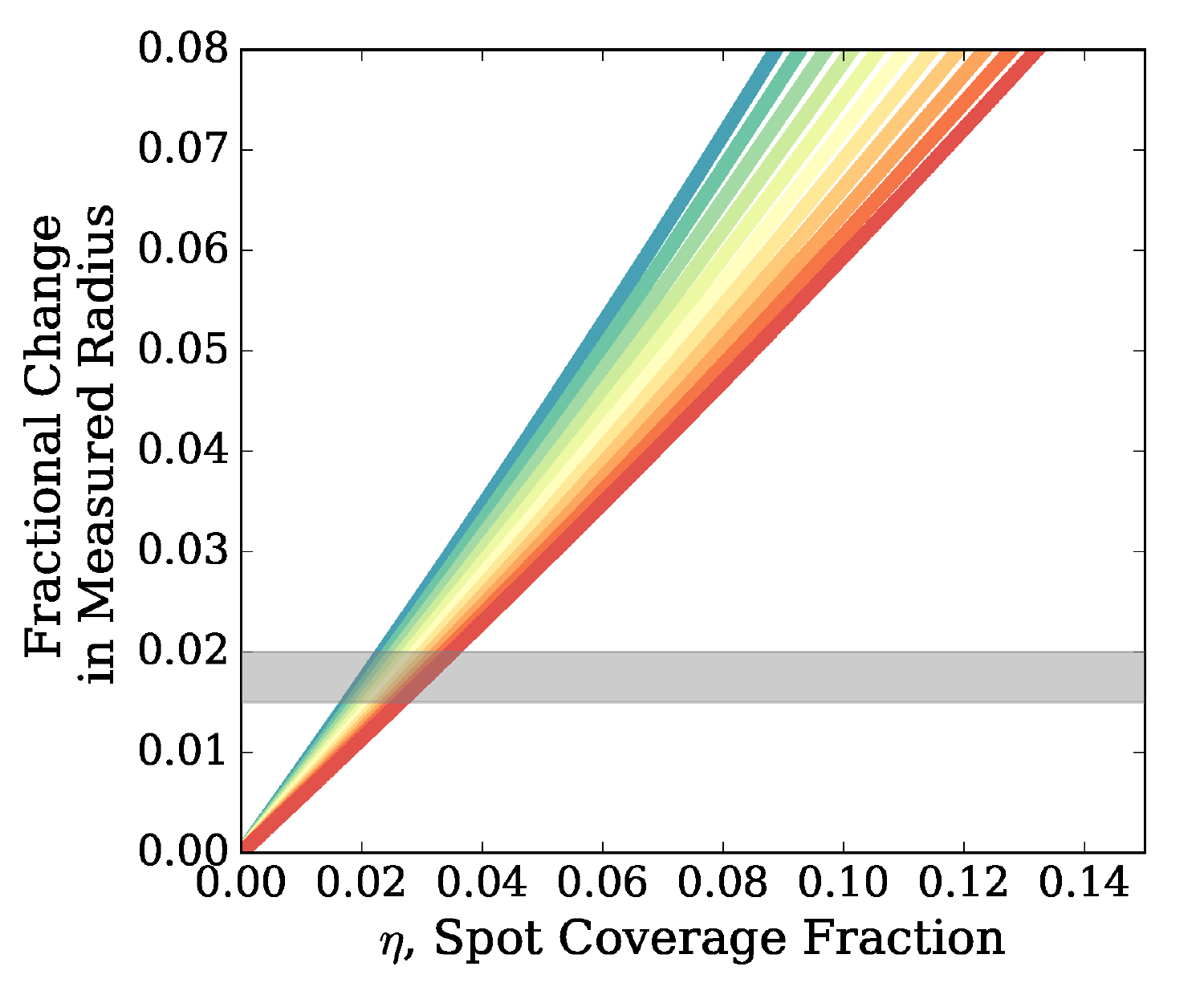}
	\caption{Using equation \ref{deltaradius}, a demonstration of how increasing spot coverage changes the measured planetary-to-stellar radius ratio for a wavelength dependent spot contrast as measured in our MCMC fits, from 4000\AA (blue) to 8000\AA (red). Over plotted in a grey band is our measured difference in radius between our two observations of WASP-52b.} \label{W52_spots}
\end{figure}

The detection of a star spot in one transit of WASP-52b suggests we must also consider the possibility of unocculted stellar features which would affect the absolute flux level of the star. Such features, if uncounted for, could mimic a Rayleigh slope, as well as introduce an offset in the absolute transit depth. We investigate the affect of unocculted star spots using the prescription given in \cite{Louden2017} where the measured transit depth, $\tilde{\delta}(\lambda)$, can be described as a function of the true depth, $\delta(\lambda)$, times a function of the spot coverage, $\eta$:
\begin{equation}
	\label{deltaradius}
	\tilde{\delta}(\lambda)=\delta(\lambda)\frac{1}{1-\eta\left(1-\frac{F_{\lambda}(spot)}{F_{\lambda}(star)}\right)}
\end{equation}
From this, we can calculate the approximate overall dimming of the star as a function of wavelength as
\begin{multline}
	F_{\lambda}(\mathrm{star,corrected})=\eta f_{\lambda} F_{\lambda}(\mathrm{star})+(1-\eta)F_{\lambda}(\mathrm{star}) \\ =[1-\eta(1-f_{\lambda})]F_{\lambda}(\mathrm{star})
\end{multline}
with $f_{\lambda}$ the spot contrast we fit with the transit light curves and $f_{\lambda}F_{\lambda}(\mathrm{star})$ equivalent to the spot's flux.  From our fit spot contrast levels, and the required spot coverage fraction to reach the observed fractional difference in transit depth (see Figure \ref{W52_spots}), we can calculate the observed stellar dimming due to spots between the two nights, and compare to photometric studies of WASP-52's variations.
\par  \cite{Hebrard2013} performed photometric monitoring of WASP-52 over two different seasons, and found the star dimmed up to  0.89\% over the course of a 16.4$\pm$0.04 day rotational period. Our observations were 42 days apart, which puts us at $\sim$2.5 periods, assuming the star presented a similar period of spot variability during our transit observations. An analysis of more recent photometric data by \cite{Louden2017} found a 17.79$\pm$0.05 day period, with a significantly higher dimming amplitude of 1.42\%. Our data requires a spot coverage of $\sim$2-3\%, corresponding to a dimming of 1.5-2\%, in agreement with previous photometric measurements.

\subsubsection{Star Spots on the Stellar Limb}
\label{limbspots}
\par
\begin{figure}
	\epsscale{1.0}
	\plotone{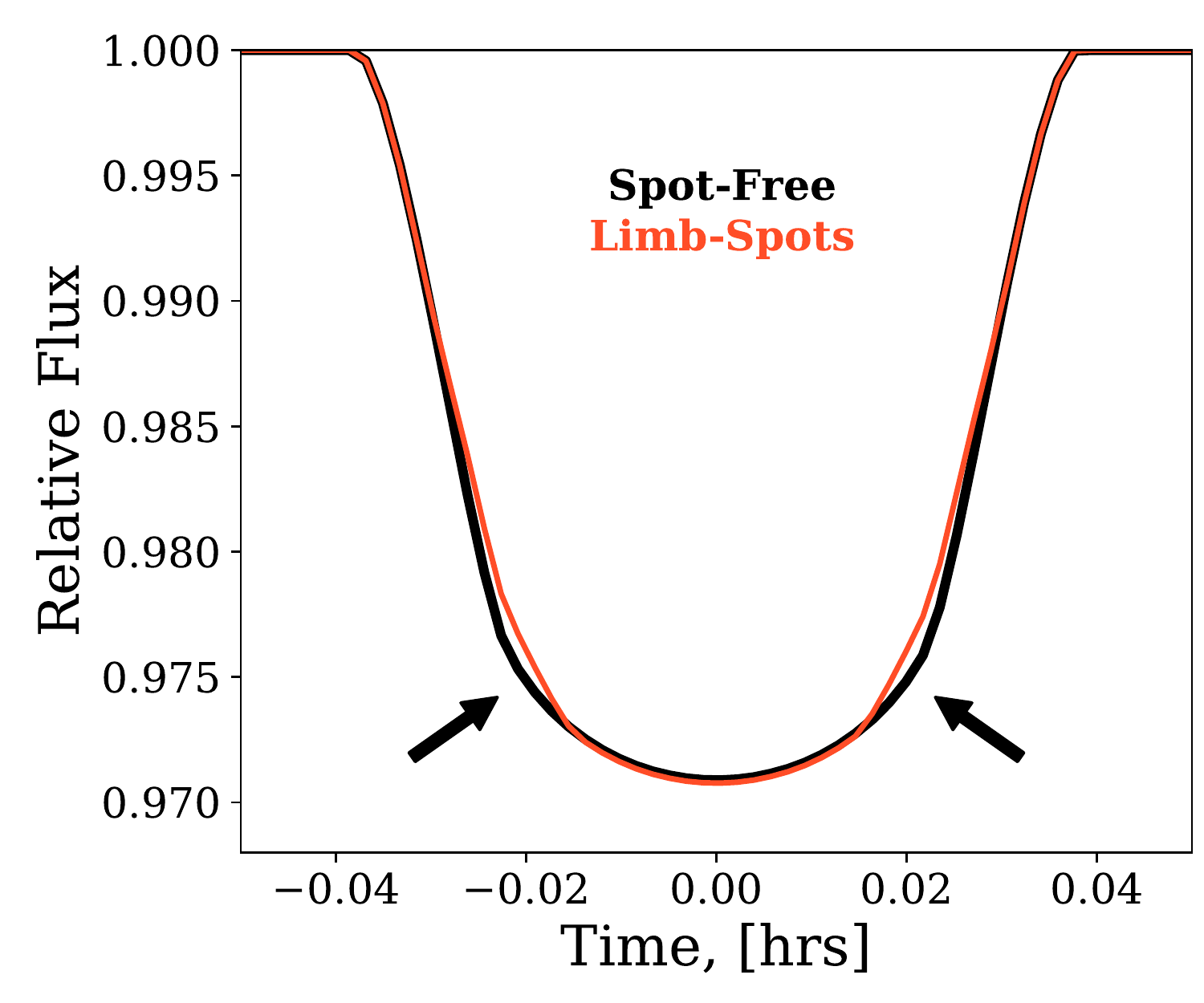}
	\caption{Effect of occulted spots on the limb on the shape of ingress/egress for a WASP-52b like planet. The black curve shows a typical transit signal for WASP-52b. In red is the resulting transit with occulted spots on the stellar limb. Note that the `typical' spot feature is not present, but instead the spotted transit simply appears to have different limb darkening.} \label{limb_lcs}
\end{figure}

 Due to our differing limb darkening measurements between our two transits of WASP-52b, we cannot ignore the possibility that occulted spots are present on the limb of the star, and manifest as variations in limb darkening. To determine whether this is a feasible explanation for our differing limb darkening measurements, we use SPOTROD to generate light curves with 2 spots on the limb of the star, in the path of the planet's transit. Both spots are identical in size, vertical location, and contrast to the occulted spot in the September 2016 transit of WASP-52b, but offset to the limb of the star. We compare these light curves to those expected in a spot-free transit (see Figure \ref{limb_lcs}) and note that spots on the limb do not appear as large `bumps' as they do during mid-transit occultations, but instead they slightly change the shape of the ingress and egress portions of transit. By eye, one does not detect the presence of a spot and therefore would not find it necessary to fit such a light curve as a spotted transit. To demonstrate how this assumption affects the measured limb darkening, we fit the transit with occulted spots on the stellar limb with a standard transit model, using the previously described BATMAN package.
 
\begin{figure}
	\epsscale{1.0}
	\plotone{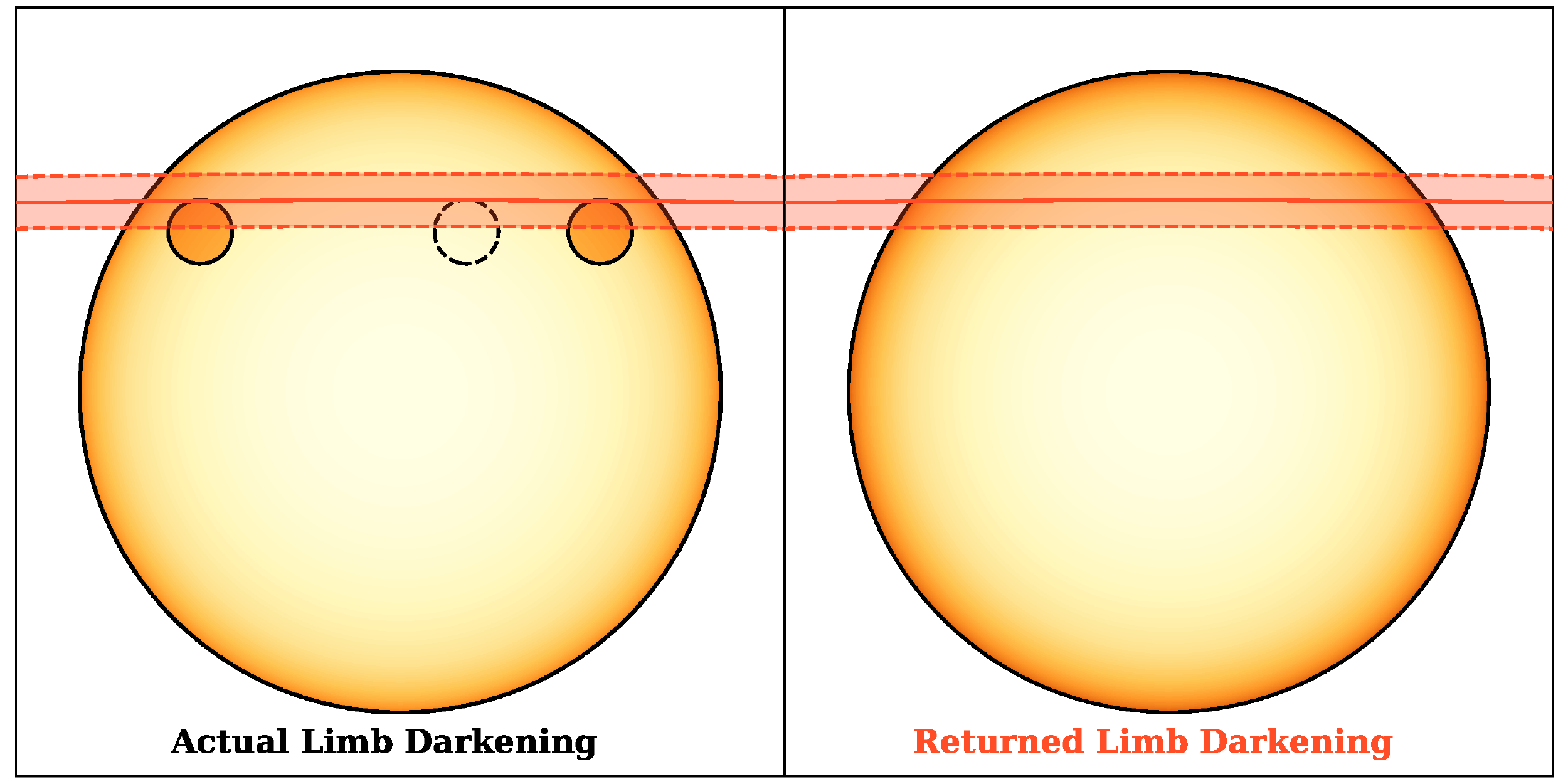}
	\caption{Effect of occulted spots on the limb on the returned limb darkening for a WASP-52b like planet. Expected limb darkening with occulted limb spots (\textbf{left}). Spots on the limb result in returned limb darkening parameters that correspond to noticeably stronger limb darkening (\textbf{right}). The shaded red region corresponds to the path of the planet during transit. The two limb spots are shown in the left panel, but not the right, because they are not assumed to be present when fitting for the observed limb darkening parameters. The dashed black circle represents the location of the measured spot in our September 2016 transit.} \label{limb_dark_vis}
\end{figure}

\begin{figure}
	\epsscale{1.0}
	\plotone{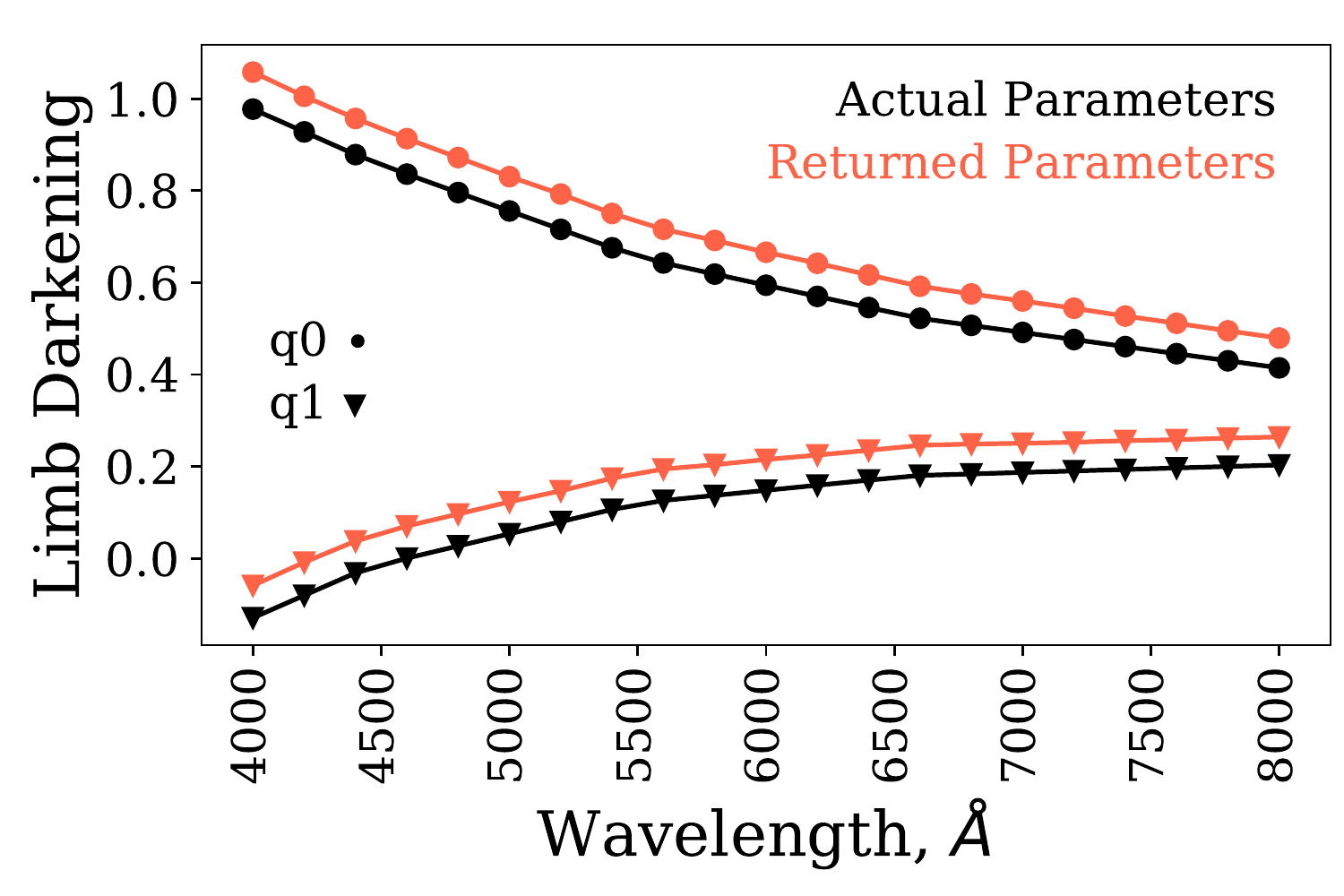}
	\hspace{0.1cm}
	\plotone{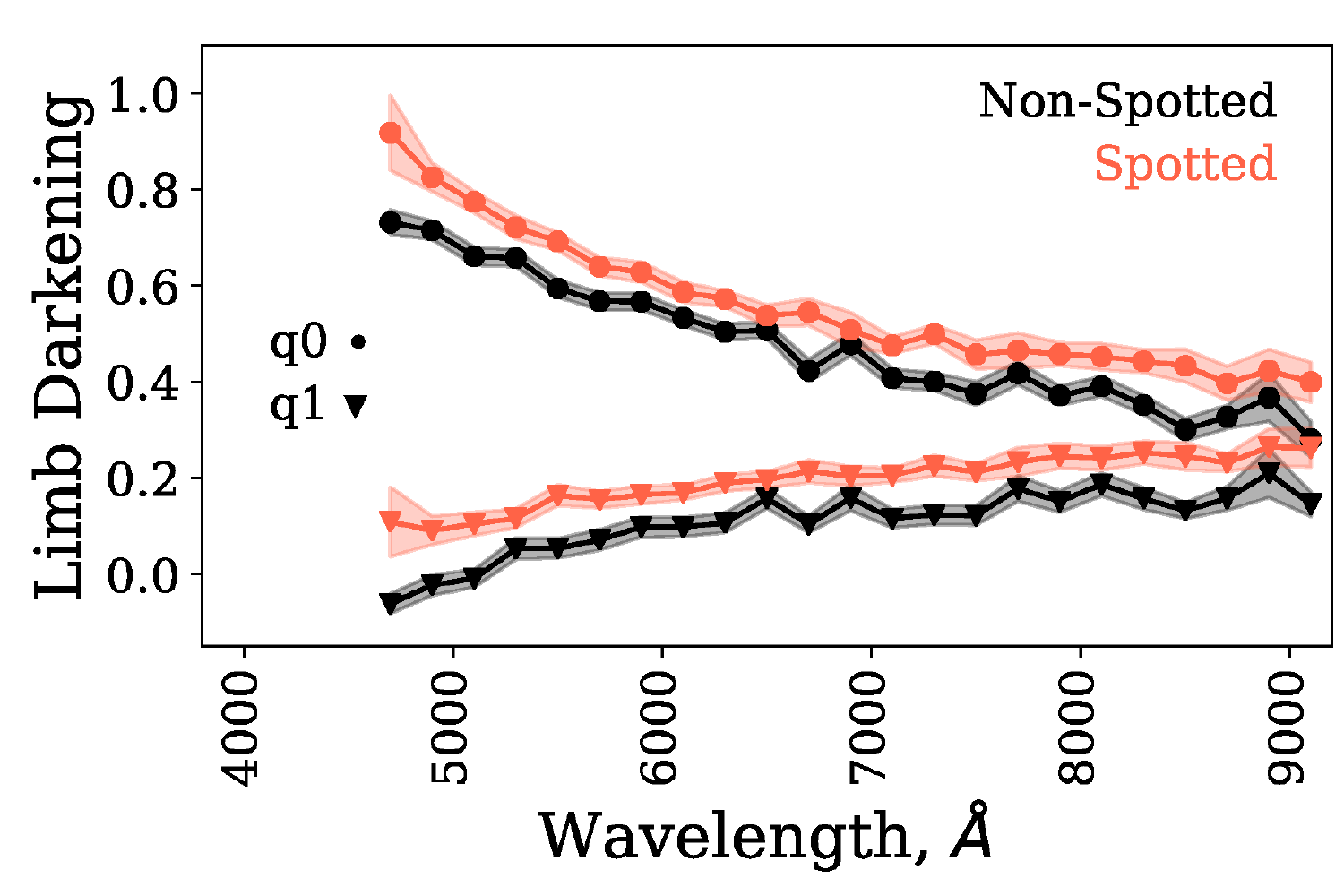}
	\caption{\textbf{TOP:} Effect of occulted spots on the limb on the returned limb darkening parameters for a WASP-52b like planet. In black are the expected, or actual, limb darkening parameters used to generate the light curve with occulted spots on the stellar limb. In red are the returned limb darkening parameters when one ignores the possibility of spots on the limb. The first quadratic parameter is denoted with circles, and the second quadratic parameter is denoted with upside-down triangles. Note that both limb darkening parameters are returned as larger values, corresponding to stronger limb darkening. \textbf{BOTTOM:} MCMC fits to both limb darkening parameters for our two nights, with the same colors and shapes as in the top plot. Note that the spotted transit is best fit with larger limb darkening values, as expected for spots on the limb.} \label{limb_dark_plt}
\end{figure}

\par As expected, the returned limb-darkening parameters do not match those used to create the light curve. Figure \ref{limb_dark_vis} shows visually the difference between the expected limb darkening with occulted spots on the limb, and the returned limb darkening. Figure \ref{limb_dark_plt} shows how the expected and returned limb darkening differ, these results are within the variations between our August and September 2016 WASP-52b transits. We find this to be a viable explanation for the differing limb darkening, though we do not attempt to fit for the presence of these limb spots, due to the degeneracy between the limb darkening and spot parameters in this part of the transit. 
\par We acknowledge that this situation is artificially contrived. However, we do note that as spots are likely to occur along the same latitude band, and we have detected a spot at this latitude in the relevant transit, it is not completely unrealistic to place additional spots in these locations. Regardless, this exercise serves only as one possible explanation for the different limb darkening measured under the circumstances that may exist for this star and to demonstrate that this possibility does produce the effect we measure.


\section{Conclusions}
\label{future}
\par We find flat spectra for both WASP-4b and WASP-52b within the errors of our measurements, implying the presence of clouds in the atmospheres of both planets. Our group is currently working on applying noise removal techniques for future data sets in order to improve our precision. We continue to schedule observations at the Magellan Baade telescope and hope to complete observations of several more targets this year.
\par With the launch of JWST in spring 2021, ground based optical studies of planetary atmospheres will remain a worthwhile use of ground-based telescope time. The NIRSPEC instrument on JWST, likely to be the most commonly used instrument for transmission spectroscopy studies, has a lower wavelength limit of 0.6$\mathrm{\mu}$m when using the prism disperser element. The prism, however, is a low resolution disperser, with a resolving power $\sim$100, comparable to our ground-based resolution. One of the major benefits of JWST is the ability to perform higher resolution transmission spectroscopy from 0.7$\mathrm{\mu}$m to 5.27$\mathrm{\mu}$m, with a resolving power up to $\sim$2700 depending on the disperser used. Therefore, ground based observations will remain important below 0.7$\mathrm{\mu}$m, both to compare measurements between 0.6$\mathrm{\mu}$m and 0.7$\mathrm{\mu}$m, and to provide transmission measurements bluewards of the lower limit of JWST. 
\par Additionally, the TESS mission is expected to find of order 1000 Sub-Neptunes in the primary mission  \citep{Sullivan2015}. As a result, we expect numerous new ground based targets for this project in the coming years. With our current methods, and improvements being made, we will be well prepared to observe and characterize these new targets.

\acknowledgements
We thank the staff at Las Campanas Observatory, without which we would be unable to carry out the observations presented in this work. This research has made use of the Exoplanet Orbit Database
and the Exoplanet Data Explorer at exoplanets.org. \nocite{exo-org} 

\facility{Magellan:Baade}
\software{\\ Astropy \citep{astropy},
\\ batman \citep{Kreidberg2015},
\\ emcee \citep{FMackey2013},
\\ Exo-Transmit \citep{Kempton2017},
\\ IPython \citep{ipython},
\\ Matplotlib \citep{matplotlib},
\\ NumPy \citep{numpy},
\\ SciPy \citep{scipy},
\\ SpectRes \citep{Carnall2017},
\\ spotrod \citep{Beky2014}}

\bibliographystyle{apj}

\appendix
\section{Check Star}
\begin{figure*}[h!]
	\epsscale{0.5}
	\plotone{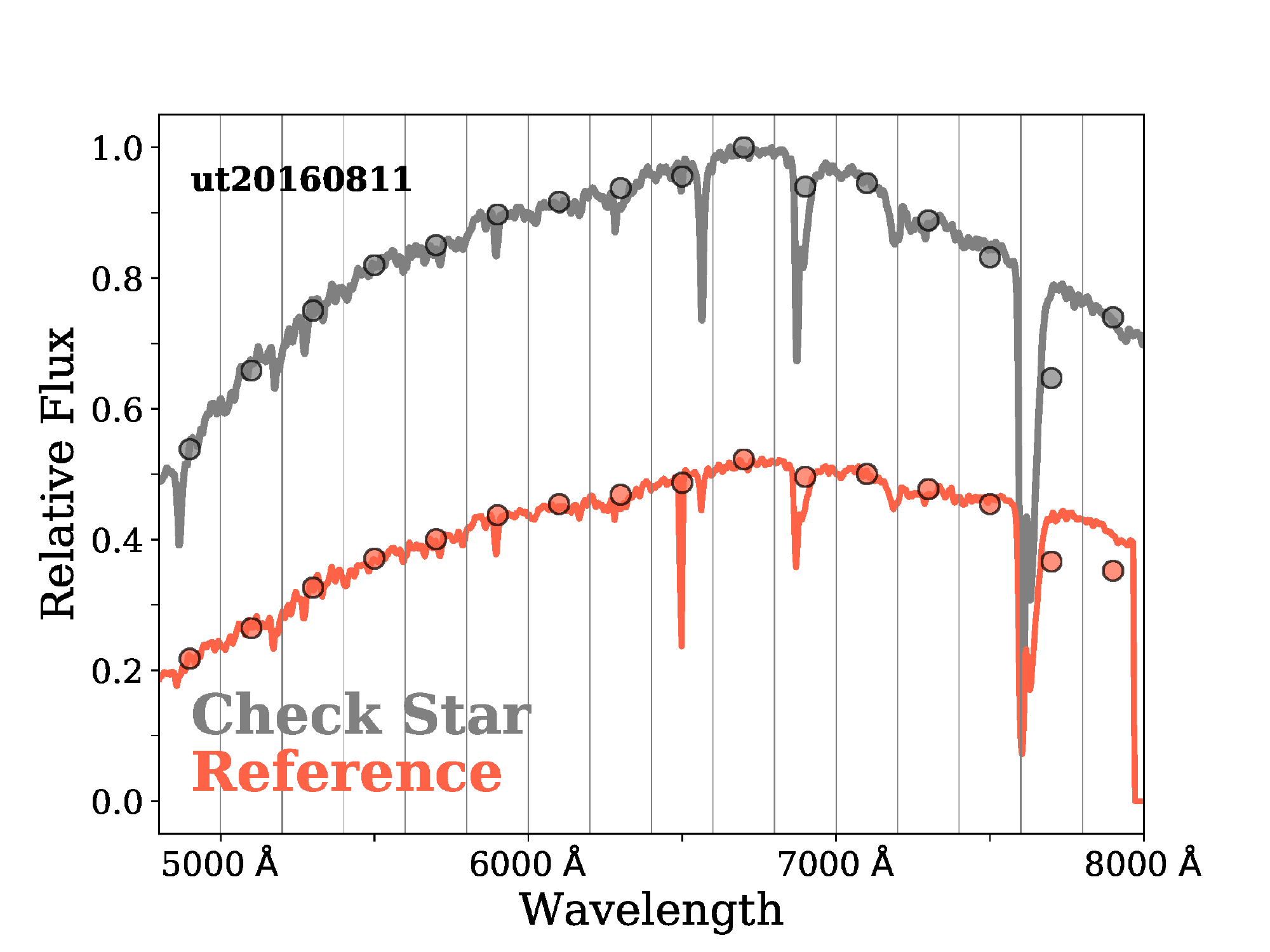}
	\caption{Spectra for our check star during our first night of WASP-52 observations from 4800 \AA through 8000 \AA. The blue line shows our check star, and the orange line shows the calibrator spectrum at one point in time. The edges of our 200 \AA binning are plotted as grey lines, with the summed values for those bins over plotted. } \label{A1}
\end{figure*}

\begin{figure*}
	\epsscale{0.8}
	\plottwo{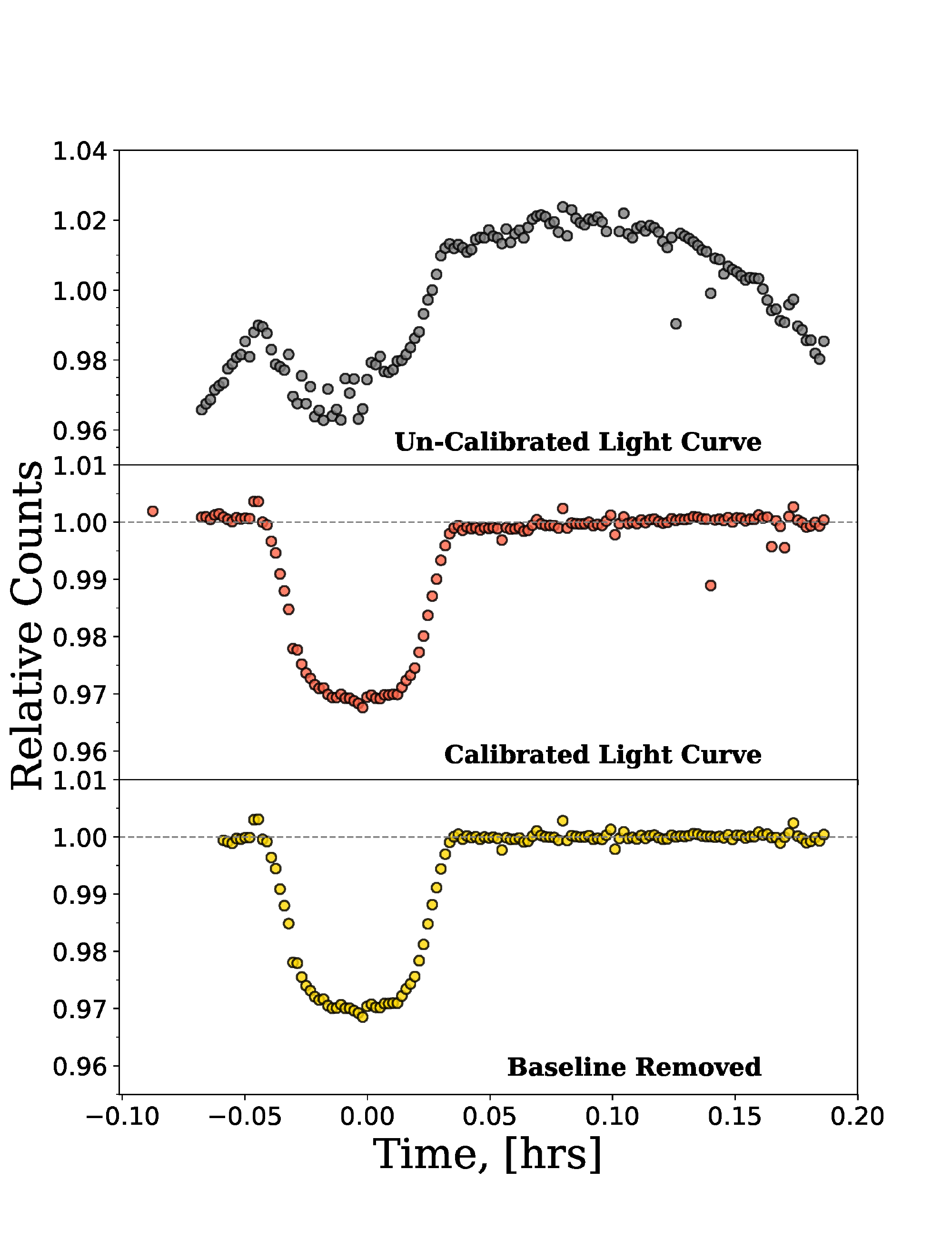}{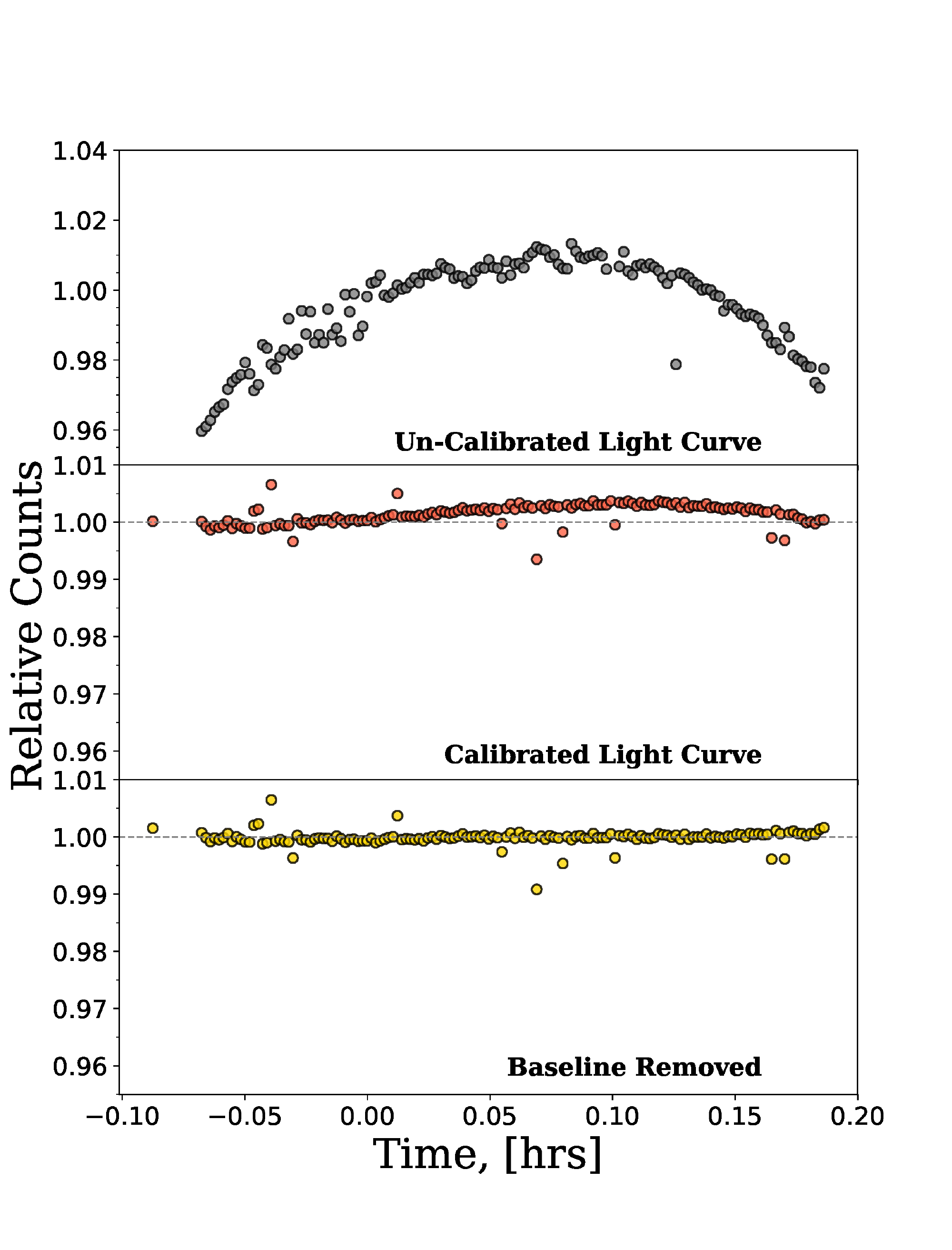}
	\caption{Showing the reduction process from the raw white light curve through the baseline-calibrated white light curve for both the target (left) and the check star (right) on the same scale. The first (top) panel of each figure is the uncalibrated, raw white light curve. The second (middle) panel shows the light curve after dividing out the combined calibrator stars. The third (bottom) panel shows the final light curve after fitting and removing a 3rd order polynomial in time to the out-of-transit baseline.} \label{A2}
\end{figure*}

As a check on our reduction process, all steps are also performed on a designated `check star' each night. Because we have a large number of simultaneously observed stars, we can easily select one as a check star to be treated as a second target instead of an additional calibrator. Here we include representative figures (see Figures \ref{A1} and \ref{A2}) for the check star from the first night of WASP-52b observations (ut20160811). 
\pagebreak
\section{Tables of Results}
\begin{deluxetable*}{cccccccc}[h!] 
\tabletypesize{\small}
\tablecolumns{5}
\tablecaption{WASP-4b: Wavelength dependent variables}
\tablewidth{0.9\textwidth}
\tablehead{
	\colhead{Center of Bin [\AA]}		&
	\colhead{R$_{p}$/R$_{star}$}		&
	\colhead{$\Delta$ R$_{p}$/R$_{star}$}	&
	\colhead{ }					&
	\colhead {q0}					&
	\colhead{$\Delta$ q0}				&
	\colhead{q1} 					&
	\colhead{$\Delta$ q1}				}
\startdata	
4700 		&	0.1560		&	0.0014		&	{ }		&	0.618		&	0.036		& 0.078	& 0.035		\\	
4900 		&	0.1590		&	0.0014		&	{ }		&	0.595		&	0.035		& 0.114	& 0.036		\\	
5100 		&	0.1530		&	0.0009		&	{ }		&	0.575		&	0.026		& 0.147	& 0.030		\\	
5300 		&	0.1571		&	0.0009		&	{ }		&	0.560		&	0.026		& 0.199	& 0.030		\\	
5500 		&	0.1529		&	0.0006		&	{ }		&	0.516		&	0.020		& 0.213	& 0.025		\\	
5700 		&	0.1534		&	0.0006		&	{ }		&	0.526		&	0.020		& 0.246	& 0.026		\\	
5900 		&	0.1557		&	0.0007		&	{ }		&	0.484		&	0.022		& 0.225	& 0.026		\\	
6100 		&	0.1557		&	0.0005		&	{ }		&	0.437		&	0.019		& 0.212	& 0.025		\\	
6300 		&	0.1570		&	0.0006		&	{ }		&	0.442		&	0.018		& 0.246	& 0.024		\\
6500 		&	0.1549		&	0.0006		&	{ }		&	0.440		&	0.021		& 0.251	& 0.025		\\
6700 		&	0.1532		&	0.0019		&	{ }		&	0.442		&	0.026		& 0.277	& 0.030		\\
6900 		&	0.1546		&	0.0015		&	{ }		&	0.430		&	0.041		& 0.278	& 0.042		\\
7100 		&	0.1559		&	0.0027		&	{ }		&	0.407		&	0.060		& 0.271	& 0.058		\\
7300 		&	0.1557		&	0.0020		&	{ }		&	0.375		&	0.052		& 0.246	& 0.054		\\
7500 		&	   ---			&	   ---			&	{ }		&	  ---			&	  ---			&   ---	&   ---		\\	
7700 		&	   ---			&	   ---			&	{ }		&	  ---			&	  ---			&   ---	&   ---		\\	
7900 		&	0.1553		&	0.0011		&	{ }		&	0.408		&	0.031		& 0.311	& 0.030		\\	
8100 		&	0.1574		&	0.0010		&	{ }		&	0.377		&	0.030		& 0.297	& 0.032		\\		
\enddata
\tablecomments{Relative planetary radius measurements and quadratic limb darkening parameters, with their corresponding errors. All results come from our MCMC fits.}
\end{deluxetable*}
\begin{deluxetable*}{cccccccc}[h!] 
\tabletypesize{\small}
\tablecolumns{5}
\tablecaption{WASP-52b: Wavelength dependent variables, ut20160811}
\tablewidth{0.9\textwidth}
\tablehead{
	\colhead{Center of Bin [\AA]}		&
	\colhead{R$_{p}$/R$_{star}$}		&
	\colhead{$\Delta$ R$_{p}$/R$_{star}$}	&
	\colhead{ }					&
	\colhead {q0}					&
	\colhead{$\Delta$ q0}				&
	\colhead{q1} 					&
	\colhead{$\Delta$ q1}				}
\startdata	
4900		&	0.1703		&	0.0015		&	{ }		&	0.715	&	0.026	&	-0.023	&	0.022	 	\\
5100		&	0.1681		&	0.0011		&	{ }		&	0.661	&	0.019	&	-0.009	&	0.018	 	\\
5300		&	0.1682		&	0.0011		&	{ }		&	0.657	&	0.019	&	0.052	&	0.022		\\
5500		&	0.1684		&	0.0010		&	{ }		&	0.594	&	0.018	&	0.054	&	0.020		\\
5700		&	0.1658		&	0.0010		&	{ }		&	0.567	&	0.018	&	0.071	&	0.021		\\
5900		&	0.1655		&	0.0009		&	{ }		&	0.566	&	0.018	&	0.098	&	0.021		\\
6100		&	0.1649		&	0.0008		&	{ }		&	0.533	&	0.017	&	0.097	&	0.020		\\
6300		&	0.1663		&	0.0008		&	{ }		&	0.504	&	0.016	&	0.107	&	0.020		\\
6500		&	0.1670		&	0.0008		&	{ }		&	0.508	&	0.017	&	0.157	&	0.021		\\
6700		&	0.1682		&	0.0013		&	{ }		&	0.423	&	0.024	&	0.103	&	0.017		\\
6900		&	0.1675		&	0.0015		&	{ }		&	0.478	&	0.024	&	0.158	&	0.026		\\
7100		&	0.1638		&	0.0012		&	{ }		&	0.407	&	0.019	&	0.115	&	0.019		\\
7300		&	0.1654		&	0.0009		&	{ }		&	0.400	&	0.019	&	0.124	&	0.020		\\
7500		&	0.1658		&	0.0013		&	{ }		&	0.375	&	0.025	&	0.122	&	0.020		\\
7700		&	0.1672		&	0.0016		&	{ }		&	0.418	&	0.024	&	0.178	&	0.027		\\
7900		&	0.1624		&	0.0013		&	{ }		&	0.371	&	0.020	&	0.151	&	0.023		\\
8100		&	0.1651		&	0.0012		&	{ }		&	0.390	&	0.021	&	0.187	&	0.024		\\
8300		&	0.1626		&	0.0012		&	{ }		&	0.352	&	0.020	&	0.157	&	0.023		\\
8500		&	0.1659		&	0.0012		&	{ }		&	0.300	&	0.023	&	0.132	&	0.017		\\
8700		&	0.1638		&	0.0015		&	{ }		&	0.327	&	0.025	&	0.157	&	0.025		\\
8900		&	0.1614		&	0.0032		&	{ }		&	0.367	&	0.050	&	0.210	&	0.050		
\enddata
\tablecomments{Relative planetary radius measurements and quadratic limb darkening parameters, with their corresponding errors. All results come from our MCMC fits.}
\end{deluxetable*}
\begin{deluxetable*}{cccccccc}[h!] 
\tabletypesize{\small}
\tablecolumns{5}
\tablecaption{WASP-52b: Wavelength dependent variables, ut20160922}
\tablewidth{0.9\textwidth}
\tablehead{
	\colhead{Center of Bin [\AA]}		&
	\colhead{R$_{p}$/R$_{star}$}		&
	\colhead{$\Delta$ R$_{p}$/R$_{star}$}	&
	\colhead{ }					&
	\colhead{q0}					&
	\colhead{$\Delta$ q0}				&
	\colhead{q1} 					&
	\colhead{$\Delta$ q1}				}
\startdata
4900		&	0.1675		&	0.0027		&	{ }		&	0.826	&	0.030	&	0.091	&	0.030	 	\\
5100		&	0.1677		&	0.0014		&	{ }		&	0.774	&	0.022	&	0.104	&	0.024	 	\\
5300		&	0.1647		&	0.0012		&	{ }		&	0.721	&	0.024	&	0.116	&	0.019		\\
5500		&	0.1616		&	0.0010		&	{ }		&	0.692	&	0.018	&	0.164	&	0.021		\\
5700		&	0.1638		&	0.0010		&	{ }		&	0.640	&	0.019	&	0.155	&	0.017		\\
5900		&	0.1656		&	0.0009		&	{ }		&	0.628	&	0.021	&	0.165	&	0.019		\\
6100		&	0.1647		&	0.0009		&	{ }		&	0.586	&	0.021	&	0.169	&	0.015		\\
6300		&	0.1650		&	0.0007		&	{ }		&	0.572	&	0.017	&	0.190	&	0.017		\\
6500		&	0.1643		&	0.0012		&	{ }		&	0.537	&	0.022	&	0.196	&	0.017		\\
6700		&	0.1659		&	0.0019		&	{ }		&	0.545	&	0.028	&	0.214	&	0.024		\\
6900		&	0.1637		&	0.0018		&	{ }		&	0.508	&	0.036	&	0.204	&	0.020		\\
7100		&	0.1670		&	0.0010		&	{ }		&	0.476	&	0.018	&	0.206	&	0.014		\\
7300		&	0.1655		&	0.0009		&	{ }		&	0.499	&	0.019	&	0.225	&	0.022		\\
7500		&	0.1706		&	0.0028		&	{ }		&	0.457	&	0.030	&	0.212	&	0.017		\\
7700		&	0.1600		&	0.0033		&	{ }		&	0.465	&	0.037	&	0.233	&	0.030		\\
7900		&	0.1641		&	0.0014		&	{ }		&	0.458	&	0.024	&	0.246	&	0.026		\\
8100		&	0.1642		&	0.0013		&	{ }		&	0.452	&	0.027	&	0.241	&	0.024		\\
8300		&	0.1651		&	0.0013		&	{ }		&	0.443	&	0.024	&	0.253	&	0.024		\\
8500		&	0.1680		&	0.0016		&	{ }		&	0.397	&	0.034	&	0.245	&	0.028		\\
8700		&	0.1668		&	0.0019		&	{ }		&	0.423	&	0.035	&	0.232	&	0.018		\\
8900		&	0.1699		&	0.0039		&	{ }		&	0.399	&	0.044	&	0.264	&	0.038		
\enddata
\tablecomments{Relative planetary radius measurements and quadratic limb darkening parameters, with their corresponding errors. All results come from our MCMC fits. Radius values have been corrected assuming a spot coverage fraction of 3\%.} 
\end{deluxetable*}
\begin{deluxetable*}{ccc}[h!]  
\tabletypesize{\small}
\tablecolumns{5}
\tablecaption{WASP-52b: Combined R$_{p}$/R$_{star}$ values}
\tablewidth{0.9\textwidth}
\tablehead{
	\colhead{Center of Bin [\AA]}		&
	\colhead{R$_{p}$/R$_{star}$}		&
	\colhead{$\Delta$ R$_{p}$/R$_{star}$}	}
\startdata	
4900		&	0.1693		&	0.0019	\\
5100		&	0.1680		&	0.0016	\\
5300		&	0.1669		&	0.0012	\\
5500		&	0.1649		&	0.0012	\\
5700		&	0.1649		&	0.0013	\\
5900		&	0.1655		&	0.0016	\\
6100		&	0.1648		&	0.0010	\\
6300		&	0.1657		&	0.0011	\\
6500		&	0.1662		&	0.0009	\\
6700		&	0.1678		&	0.0011	\\
6900		&	0.1663		&	0.0012	\\
7100		&	0.1661		&	0.0013	\\
7300		&	0.1654		&	0.0016	\\
7500		&	0.1662		&	0.0011	\\
7700		&	0.1660		&	0.0012	\\
7900		&	0.1634		&	0.0012	\\
8100		&	0.1648		&	0.0011	\\
8300		&	0.1638		&	0.0011	\\
8500		&	0.1666		&	0.0011	\\
8700		&	0.1647		&	0.0012	\\
8900		&	0.1625		&	0.0030	
\enddata
\tablecomments{Relative planetary radius measurements and quadratic limb darkening parameters, with their corresponding errors. All results come from our MCMC fits.}
\end{deluxetable*}
\end{document}